%% file: arxiv_main.tex
\documentclass[sigconf,nonacm]{acmart}

\setcopyright{none}
\renewcommand\footnotetextcopyrightpermission[1]{}

\usepackage{booktabs}
\usepackage[table]{xcolor}
\usepackage{graphicx}
\usepackage{subcaption}
\usepackage{adjustbox}
\usepackage{placeins}
\usepackage{microtype}
\usepackage{seqsplit}
\usepackage{amsmath}
\usepackage{enumitem}
\usepackage{pifont}

\AtBeginDocument{%
}

\DeclareRobustCommand{\code}[1]{\texttt{\seqsplit{#1}}}
\DeclareRobustCommand{\labeltext}[1]{\ifmmode\text{\sffamily #1}\else\textsf{#1}\fi}
\newcolumntype{L}[1]{>{\raggedright\arraybackslash}p{#1}}
\newcolumntype{C}[1]{>{\centering\arraybackslash}m{#1}}
\newcommand{\capYes}{\ding{51}}
\newcommand{\capPart}{$\triangle$}
\newcommand{\capNo}{--}

\title{Evidence-in-the-Loop: Trace-Driven Optimization for Customer-Service LLM Agents}

\author{Chunming Wu}
\email{mingo.wu@flsdex.com}
\affiliation{%
  \city{Shenzhen}
  \country{China}
}

\author{Dafei Qiu}
\email{david.qiu@flsdex.com}
\affiliation{%
  \city{Shenzhen}
  \country{China}
}

\author{Charles Quan}
\email{charles.quan@flsdex.com}
\affiliation{%
  \city{Shenzhen}
  \country{China}
}

\author{Jun Wu}
\authornote{Corresponding author.}
\email{john.wu@flsdex.com}
\affiliation{%
  \city{Shenzhen}
  \country{China}
}

\author{Suipeng Li}
\email{patrick.s.li@flsdex.com}
\affiliation{%
  \city{Shenzhen}
  \country{China}
}

\author{Mo Wu}
\email{morgan.mo@flsdex.com}
\affiliation{%
  \city{Shenzhen}
  \country{China}
}

\author{Gavin Xie}
\email{gavin.xie@flsdex.com}
\affiliation{%
  \city{Shenzhen}
  \country{China}
}

\author{Hope Chen}
\email{hope.chen@flsdex.com}
\affiliation{%
  \city{Shenzhen}
  \country{China}
}

\author{Max Yao}
\email{max.yao@flsdex.com}
\affiliation{%
  \city{Shenzhen}
  \country{China}
}

\author{Congde Yuan}
\email{congdeyuan@gmail.com}
\affiliation{%
  \city{Shenzhen}
  \country{China}
}

\begin{document}

\begin{abstract}
Production customer-service bots must improve answer quality across iterative releases, yet large language models must not bypass evidence boundaries, policy rules, or human-handoff safeguards. We present an \textbf{Evidence-Grounded Customer-Service Agent Workflow} deployed in a real-world customer-service setting. BM25 recall, issue-title-vector recall, issue-description-vector recall, weighted RRF fusion, and cross-encoder reranking construct grounded FAQ evidence for controlled LLM decisions. Policy-guided orchestration then combines this RAG evidence with scenario-specific rule evidence, conversation memory, and clarification state inside a fixed LangGraph DAG~\cite{langgraph2024}. The paper contributes three reusable deployment patterns: \textbf{hybrid RAG evidence construction}, where multi-channel retrieval and reranking produce auditable FAQ candidates; \textbf{evidence-grounded issue/action decision}, where an Evidence-Grounded Decision Module selects an issue/action from typed FAQ evidence and scenario-specific rule evidence; and \textbf{trace-driven RAG and reranker improvement}, where traces diagnose whether failures come from recall, ranking, final candidate selection, clarification, rule-derived evidence, or action policy, and where reranker fine-tuning is evaluated not only for in-domain gain but also for forgetting risk.
The evaluation is organized as a Diagnosis $\rightarrow$ Optimization $\rightarrow$ Validation story rather than a single leaderboard number. In E1, a 309-query non-test diagnostic subset from the curated historical T-Set sample shows that hybrid recall provides high top-50 coverage (96.76\%), but the base BGE reranker reaches only 56.31\% Hit@1; under this diagnostic pipeline, replacing Qwen3.5-27B with GPT-4o changes final issue-election accuracy by less than 1pp, suggesting that backbone scaling alone is not the dominant bottleneck. In E2, the curated historical T-Set sample (605 anonymized conversation cases / 1,427 turns) provides a cleaned FAQ subset: 581 turn-level queries labeled with gold FAQ issue IDs, split into 492 train, 52 development, and 37 held-out test queries. The 544 non-test queries show that RRF fusion reaches 99.3\% Hit@50 and the base reranker reaches 95.2\% Hit@10, while teacher-score distillation improves held-out reranker Hit@1 from 56.76\% to 75.68\% and improves the six-task C-MTEB average from 66.09 to 66.58. In E3, on a sealed 200-session B-Set, the base and distilled rerankers share the same RRF candidate pool, while the distilled reranker improves issue-turn Rerank@1 from 69.4\% to 79.2\% and raises KB-grounded session accuracy from 86.5\% to 88.5\%. In E4, with the base reranker fixed, a DPO-trained final-decision LLM raises KB-grounded B-Set session accuracy to 90.5\%, isolating decision-stage post-training from reranker changes. A post-launch operational review on two equally sized production samples reports 89.52\% reviewed accuracy for the Agent workflow and 79.00\% for the legacy RAG-only workflow under the same session-level rubric. The main takeaway is that production customer-service agents should first diagnose where the decision pipeline fails---retrieval, reranking, or final selection---and then apply the right update to the evidence layer or the final decision layer, rather than relying solely on a larger or more autonomous backbone model.
\end{abstract}


\maketitle

\section{Introduction}
Production customer-service agents face two closely related challenges. First, user intents, product policies, and support knowledge evolve continuously, while the base LLM and existing FAQ corpus may lag behind these changes. Second, when evidence is incomplete, relying on the LLM alone can lead to hallucinated, over-generalized, or policy-inconsistent responses. In customer-service workflows, such deviations may mislead users, increase complaint risk, and create operational or compliance exposure. We therefore avoid treating the LLM as a free-form answer generator. Instead, the system supplies the LLM with current FAQ evidence, rule-derived scenario evidence, and replay-verified feedback, so that responses are grounded in auditable evidence and can be continuously reviewed and improved.

Customer-service assistants in regulated, high-stakes domains combine heterogeneous evidence sources: unstructured FAQ articles, permissioned context fields, and deterministic business rules. Existing systems typically adopt one of two extremes:
\begin{itemize}[leftmargin=1.2em]
  \item \textbf{Pure RAG pipelines}~\cite{lewis2020rag,gao2024ragsurvey} retrieve semantically similar FAQ entries and generate answers with an LLM. They handle paraphrase and long-tail wording well, but may ignore scenario-specific context fields and strict policy constraints when structured evidence is available.
  \item \textbf{Hard rule-first pipelines}~\cite{young2013pomdp,garcez2023neurosymbolic} invoke backend services and return rule-matched answers directly. They enforce policy precisely when triggers fire, but over-trigger on noisy context fields and cannot reconcile conflicting or incomplete evidence.
\end{itemize}

We study a deployed \emph{Evidence-Grounded Customer-Service Agent Workflow}. It lets the LLM interpret intent and select an answer, clarification, or handoff from normalized FAQ evidence, rule-derived scenario evidence, conversation memory, and clarification state, while external calls, rule execution, and serving boundaries remain under workflow control. We use \emph{Evidence-in-the-Loop} for the surrounding trace-driven process in which the same evidence localizes failures and guides KB, reranker, prompt/rule, or final-decision-model updates.

The implementation uses a LangGraph~\cite{langgraph2024} \code{StateGraph} organized as a fixed DAG with session analysis, evidence availability checks, intent routing, evidence retrieval, and final issue/action selection. Unlike an open-ended tool loop~\cite{yao2023react,singh2025agenticrag}, retrieval, the MCP context adapter, and rules only provide evidence; Python code enforces hard boundaries such as evidence isolation, clarification limits, and human-handoff rules; and the final LLM selects from the fused evidence under these constraints.

\paragraph{Contributions.}
\label{contrib:controller}
\label{contrib:ranking}
Our contributions to applied customer-service LLM systems are:
\begin{itemize}[leftmargin=1.2em]
  \item \textbf{Hybrid RAG evidence construction.} We report a deployed retrieval pipeline that combines lexical and semantic FAQ recall, RRF fusion, and cross-encoder reranking to produce auditable FAQ evidence for customer-service decisions.
  \item \textbf{Evidence-grounded issue/action decision.} We show how FAQ evidence, rule-derived scenario evidence, memory, and clarification state can be fused before the final LLM chooses whether to answer, clarify, or hand off, while deterministic code enforces serving boundaries.
  \item \textbf{Trace-driven improvement.} We introduce a replay-based review process that identifies whether a failure comes from recall, reranking, final selection, clarification, or policy, and turns reviewed failures into KB updates, reranker hard negatives, decision preference pairs, or rule/prompt fixes.
\end{itemize}

\begin{table}[t]
  \caption{Reusable workflow patterns in the Evidence-Grounded Customer-Service Agent Workflow.}
  \label{tab:design-vs-eng}
  \centering
  \footnotesize
  \setlength{\tabcolsep}{2pt}
  \begin{tabular*}{\columnwidth}{@{\extracolsep{\fill}}>{\raggedright\arraybackslash}p{0.28\columnwidth}>{\raggedright\arraybackslash}p{0.34\columnwidth}>{\raggedright\arraybackslash}p{0.32\columnwidth}@{}}
    \toprule
    \textbf{Pattern} & \textbf{Role} & \textbf{Production support} \\
    \midrule
    Hybrid RAG evidence construction &
      Build auditable FAQ candidates from lexical and semantic retrieval. &
      BM25 + vector recall; weighted RRF; cross-encoder reranking \\
    Evidence-grounded issue/action decision &
      Select answer, clarification, or handoff from fused evidence. &
      Fixed LangGraph DAG; rule-derived evidence; clarification state \\
    Trace-driven improvement &
      Turn replay failures into targeted updates. &
      Replay traces; Human+AI review; hard negatives and preference pairs; sealed B-Set validation \\
    \bottomrule
  \end{tabular*}
\end{table}

\begin{figure*}[t]
  \centering
  \includegraphics[width=0.86\textwidth]{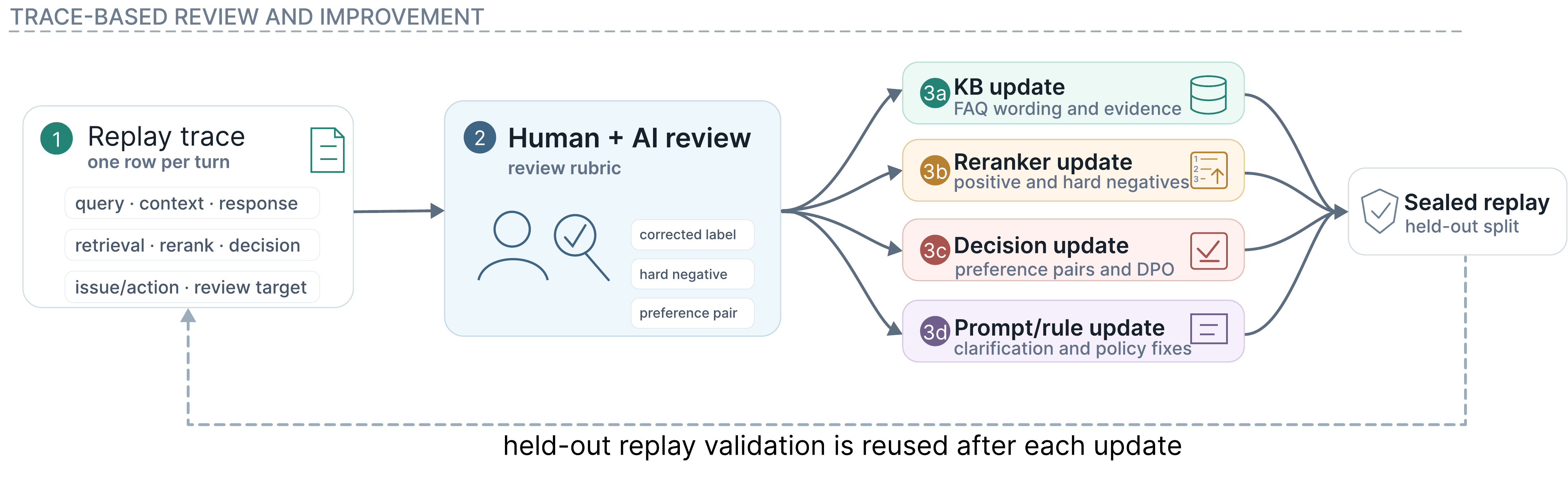}
  \caption{Trace-driven review and improvement loop. Batch replay surfaces failures with retrieval evidence; Human+AI review converts them into corrected labels, preference pairs, and hard negatives; KB, reranker, decision-prompt, or rule updates are validated on held-out replay validation splits before integration.}
  \label{fig:trace_review_loop}
\end{figure*}

Figure~\ref{fig:trace_review_loop} summarizes the replay artifacts, review outputs, and update paths used by this process.

\paragraph{What we evaluate now vs.\ what the system supports.}
Section~\ref{sec:experiments} evaluates evidence recall, reranking, final selection, and session-level outcomes. It does not claim complete coverage of MCP/rule evidence, action behavior, or downstream business outcomes.

\section{Related Work}
RAG grounds LLM answers in retrieved documents~\cite{lewis2020rag,gao2024ragsurvey}; adaptive variants decide when to retrieve or critique passages~\cite{asai2024selfrag,jeong2024adaptiverag,yan2024crag}, and customer-service RAG has been deployed with structured knowledge such as ticket knowledge graphs~\cite{xu2024cskg}. Hybrid lexical--semantic retrieval and reranking are standard ways to improve candidate quality~\cite{robertson2009bm25,chen2022hybrid,bruch2023fusion,nogueira2019bert,xiao2024bge}; Azure AI Search exposes a similar BM25/vector/RRF/semantic-ranker stack~\cite{azureHybridSearch2026,azureHybridRRF2026,azureSemanticRanker2026}. Our system uses retrieval and reranking as one evidence channel rather than the whole control surface.

Tool-augmented agents such as ReAct, Toolformer, and ToolLLM interleave reasoning with API calls~\cite{yao2023react,schick2023toolformer,qin2023toolllm}, while task-oriented dialogue and constrained routers separate state, slots, and action policy~\cite{young2013pomdp,wen2017network,chung2023instructtods,sun2024dfarag}. We adopt the production constraint that the LLM may route intent, ask, and select an issue, but policy-guided orchestration governs MCP/rule-evidence access, clarification bounds, and human handoff. This also aligns with work arguing that practical LLM systems improve through explicit memory, protocols, and feedback harnesses rather than model weights alone~\cite{shinn2023reflexion,zhou2026externalization,zhao2025agentloop}.

\begin{table}[t]
  \caption{Capability positioning against common customer-service architectures. The table summarizes typical architectural support rather than benchmark performance.}
  \label{tab:positioning}
  \centering
  \footnotesize
  \setlength{\tabcolsep}{1.5pt}
  \renewcommand{\arraystretch}{1.18}
  \begin{tabular*}{\columnwidth}{@{\extracolsep{\fill}}L{0.47\columnwidth}C{0.16\columnwidth}C{0.15\columnwidth}C{0.12\columnwidth}@{}}
    \toprule
    \textbf{Capability} & \textbf{RAG-only} & \textbf{Rule-first} & \textbf{Ours} \\
    \midrule
    Intent-aware routing & \capPart & \capYes & \capYes \\
    Hybrid FAQ + rule-evidence fusion & \capPart & \capNo & \capYes \\
    Rule evidence in LLM decision & \capNo & \capPart & \capYes \\
    Bounded issue/action decision & \capPart & \capYes & \capYes \\
    Stateful clarify / handoff & \capPart & \capYes & \capYes \\
    Replay-based improvement & \capPart & \capPart & \capYes \\
    \bottomrule
  \end{tabular*}
  \vspace{2pt}
  \begin{minipage}{0.96\columnwidth}
    \scriptsize \ding{51}: native / typical support; $\triangle$: possible with add-on integration; --: not typical.
  \end{minipage}
\end{table}

Table~\ref{tab:positioning} summarizes how the Evidence-Grounded Customer-Service Agent Workflow combines retrieval, rule-derived evidence, bounded decisions, and replay-based improvement.

\section{Problem Setting}
Given a user request $q$, conversation history $H$, and lightweight session context $u$ (such as evidence availability and dialogue state), the system predicts a response action $a$ and response content $y$.
The deployed system stores conversation history across turns. Serving prompts use a bounded history window, while the retained history supports context recovery, audit, and offline trace-driven improvement.

Actions follow task-oriented dialogue conventions~\cite{young2013pomdp,wen2017network}:
\begin{itemize}[leftmargin=1.2em]
  \item \labeltext{Answer}: return a grounded response from a selected FAQ issue,
  \item \labeltext{Clarify}: ask for necessary context when evidence is insufficient,
  \item \labeltext{Human Handoff}: hand off to a human agent.
\end{itemize}
Non-service and policy-blocked inputs are handled by intent-routing guardrails before FAQ selection; they are not the focus of the FAQ issue/action evaluation.

The system must jointly satisfy:
\begin{itemize}[leftmargin=1.2em]
  \item \textbf{Factual correctness} grounded in knowledge-base or backend evidence,
  \item \textbf{Business-policy compliance} via rule grounding without over-triggering,
  \item \textbf{Multi-turn robustness} when user context is incomplete,
  \item \textbf{Evidence robustness} when only part of the evidence set is available.
\end{itemize}

\section{Method}
\subsection{Notation and Objectives}
\label{sec:notation}
Let $q_t$ denote the latest user utterance at turn $t$, $H$ the trimmed dialogue history, and $u$ lightweight user/session state such as evidence availability and dialogue state. The intent-routing LLM outputs a structured tuple
\begin{equation}
\label{eq:classify}
  \Gamma(q_t, H, u) = \bigl(s_t,\; \eta_t,\; \tilde{q}_t,\; a_{\text{route},t}\bigr),
\end{equation}
where $s_t \in \mathcal{S}$ is an intent/scene label ($|\mathcal{S}|{=}7$ in production), $\eta_t$ denotes context needs used by the rule-derived evidence branch, $\tilde{q}_t$ a rewritten retrieval query, and $a_{\text{route},t}$ a routing signal such as a policy guardrail or a state-qualified handoff cue. The decision-stage LLM maps a fused evidence pool $\mathcal{C}$ to
\begin{equation}
\label{eq:decide}
  \Delta(\mathcal{C}, q_t, H) = \bigl(\hat{\iota},\; \hat{a}\bigr),
\end{equation}
with $\hat{\iota}$ an elected FAQ issue ID and $\hat{a}$ a dialogue action. The final response text is produced by an audited response stage conditioned on the selected issue/action and supporting evidence. Our goal is to maximize grounded issue/action accuracy while preserving hard guardrails (evidence-boundary isolation, clarification limits, and human handoff) independent of $\Delta$.

\subsection{Evidence-Grounded Workflow Design}
\label{sec:llm-centric}
The framework is an evidence-grounded workflow implemented as a LangGraph~\cite{langgraph2024} \code{StateGraph}. Each non-terminal customer-service turn invokes the LLM only through bounded workflow stages. The logical workflow is:
\begin{center}
  \footnotesize
  Session analysis $\rightarrow$
  Intent-routing LLM $\rightarrow$
  \{rule-derived evidence path \textbar{}\textbar{} RAG evidence path\}
  $\rightarrow$ evidence fusion $\rightarrow$ final issue/action selection
\end{center}
\begin{itemize}[leftmargin=1.2em]
  \item \textbf{Intent-routing LLM}: intent and scene routing over a deployment-defined label set, context extraction, and routing-signal judgment.
  \item \textbf{Guardrail routing and parallel evidence paths}: intent routing surfaces guardrail routes before FAQ selection. Policy-blocked inputs can be resolved immediately; handoff remains state-qualified and is governed by dialogue state. Otherwise, the workflow uses context needs to access permissioned context fields through MCP and construct rule-derived evidence in parallel with RAG evidence from query rewriting, hybrid FAQ retrieval, weighted RRF fusion, and reranking.
  \item \textbf{Decision-stage LLM}: issue/action selection over typed evidence that combines general FAQ candidates with scenario-specific rule evidence, yielding an issue ID, \labeltext{Clarify}, or \labeltext{Human Handoff}. An optional formatting stage adds a collapsible KB source to the final answer.
\end{itemize}
In the base workflow, both stages use a shared LLM through customer-service-specific prompts and structured output contracts; Section~\ref{sec:exp-bvolume} separately evaluates a DPO-trained final-decision LLM as a post-training variant.
Prompts are versioned and served through the production prompt infrastructure.
Non-LLM modules---hybrid retrieval, MCP context access, and rule matching---supply evidence rather than dictating the final outcome.

\begin{figure}[t]
  \centering
  \includegraphics[width=\columnwidth]{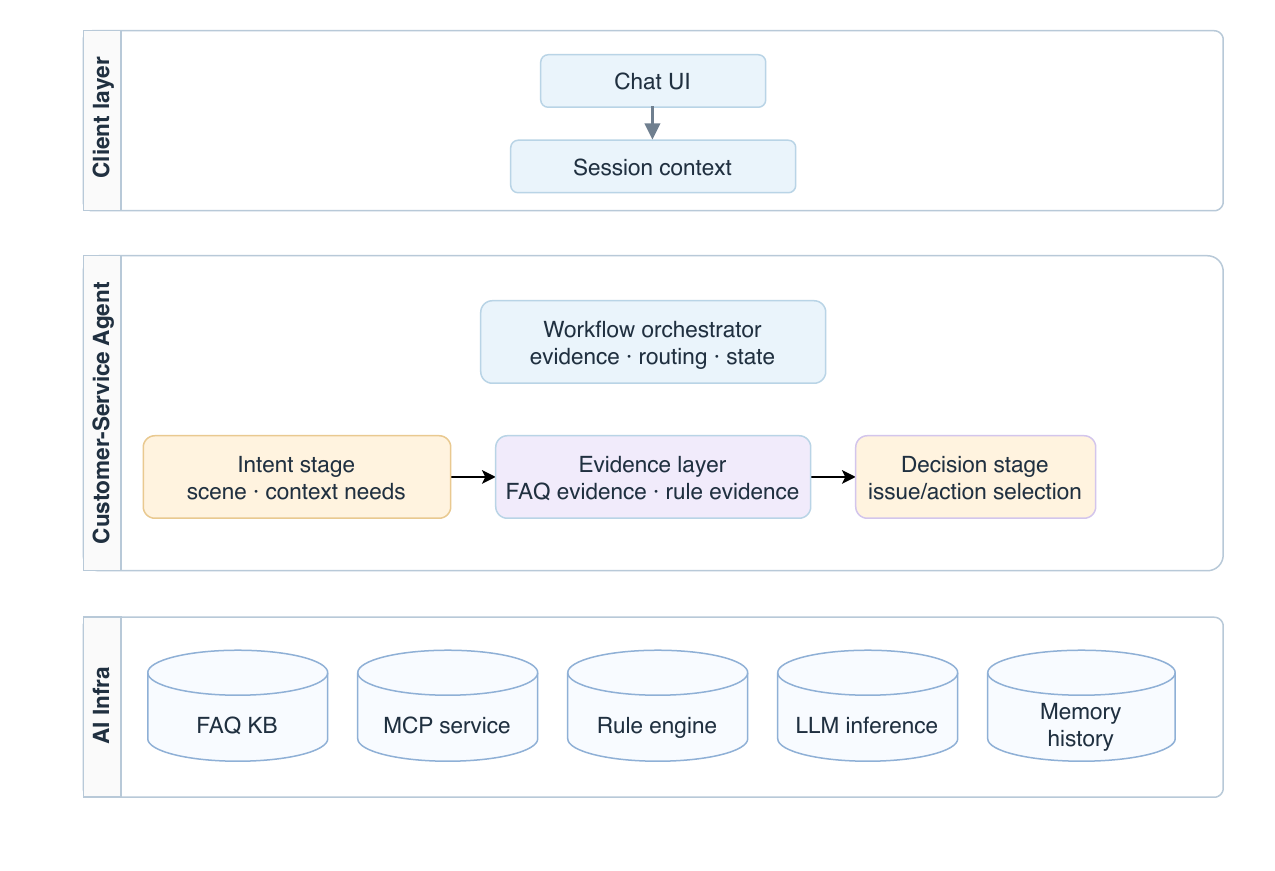}
  \caption{Three-layer logical architecture: client channel, Evidence-Grounded Customer-Service Agent Workflow, and shared AI infrastructure with persistent conversation memory.}
  \label{fig:system_architecture}
\end{figure}

Figure~\ref{fig:system_architecture} situates the workflow modules across the production architecture, including the conversation-memory store that preserves long-horizon interaction history outside the prompt window.

\begin{figure}[t]
  \centering
  \includegraphics[width=\columnwidth]{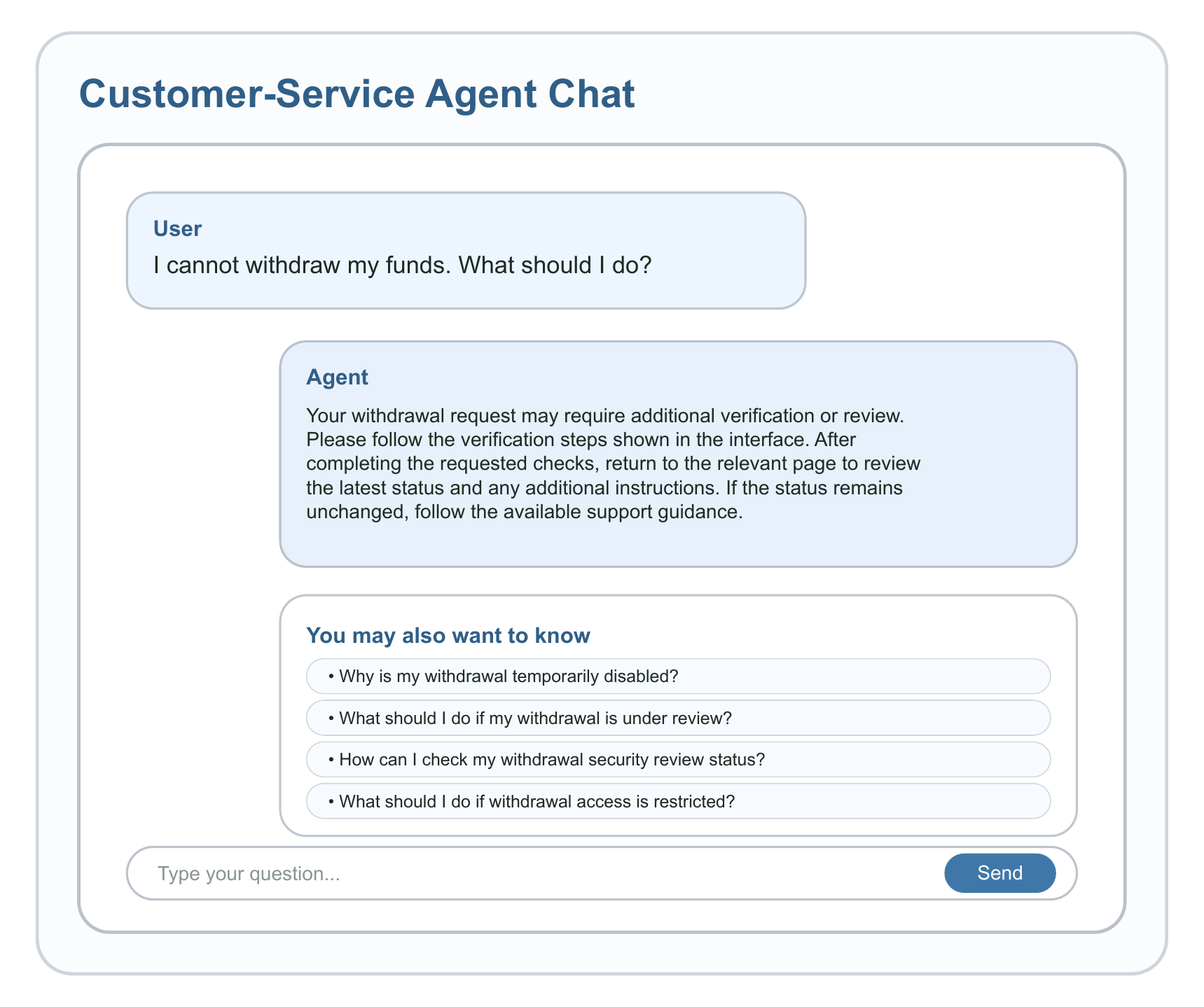}
  \caption{User-facing customer-service chat surface. The user sees a grounded answer and related question suggestions in a bot-style dialog; internal FAQ issue IDs, rule evidence, and retrieval scores are not exposed.}
  \label{fig:agent_response_interface}
\end{figure}

Figure~\ref{fig:agent_response_interface} illustrates the presentation layer of the same evidence pipeline: the final answer is shown as natural language, while neighboring FAQ candidates can be rendered as related questions for adjacent self-service paths.

\subsection{Workflow Overview}
\label{sec:workflow}
The session-analysis stage reads persisted session memory and trims dialogue history to configured bounds ($H_{\max}$ turns and $L_{\max}$ characters).
The turn first enters the intent-routing LLM, which produces intent, context needs, rewritten-query, and routing signals.
Depending on those signals, the workflow either resolves a guarded path under deterministic serving conditions or constructs two kinds of evidence in parallel: rule-derived evidence from MCP-accessed context fields and rule-engine matches, and general FAQ evidence from hybrid retrieval and reranking.
For ordinary FAQ turns, scenario-specific rule evidence and general FAQ candidates converge on the final issue/action selection stage.

\begin{figure}[t]
  \centering
  \includegraphics[width=\columnwidth]{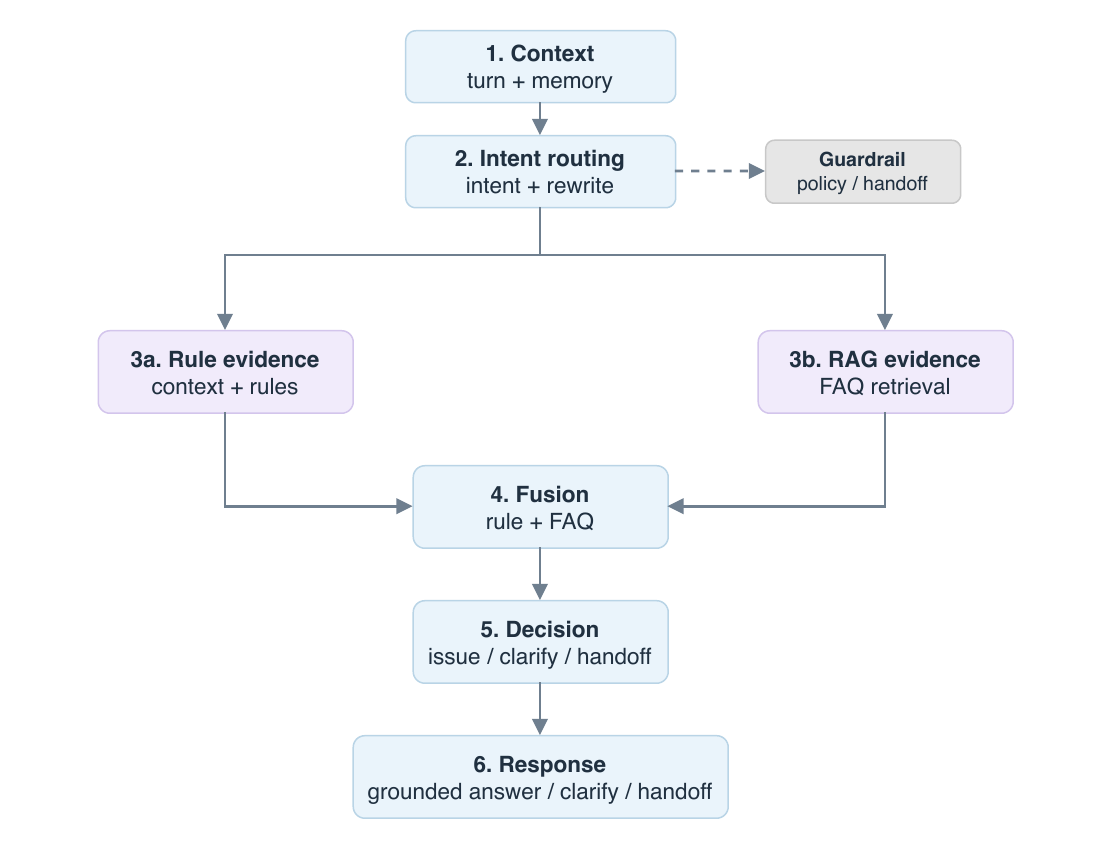}
  \caption{Conceptual workflow from context construction through intent routing, evidence construction, evidence-grounded decision making, and audited response. Guardrail outcomes may bypass issue election; otherwise rule and RAG evidence are built in parallel, fused, and audited by the decision stage.}
  \label{fig:workflow_pipeline}
\end{figure}

Figure~\ref{fig:workflow_pipeline} summarizes the workflow node sequence: after intent routing, state-qualified guardrail paths may be resolved before FAQ issue selection, while rule-derived evidence and general FAQ evidence are constructed in parallel and then converge before the final decision LLM. The internal RAG path is expanded separately in Figure~\ref{fig:retrieval_pipeline}. The DAG enables per-node logging and partial-evidence preservation.

\subsection{Evidence-Conditioned Workflow Routing}
\label{sec:routing}
The intent-routing node implements Equation~\eqref{eq:classify}.
Let $\mathcal{A}_{\text{guard}}$ denote routing signals that may be resolved before FAQ evidence construction, such as policy-blocked inputs or state-qualified handoff.
The routing function is deterministic in evidence availability and $a_{\text{route},t}$:
\begin{equation}
\label{eq:route}
  \operatorname{next}(u, a_{\text{route},t}) =
  \begin{cases}
    \labeltext{End}, & a_{\text{route},t}\in\mathcal{A}_{\text{guard}},\\
    \labeltext{Parallel}, & \text{otherwise}.
  \end{cases}
\end{equation}
\labeltext{Parallel} denotes the shared evidence-fusion path for ordinary customer-service turns.
FAQ recall runs in the RAG path, while the rule-derived evidence path reads context fields through MCP and passes matched fields to the rule engine.
If no rule condition applies, the rule-derived evidence set is empty while FAQ recall still runs.
Programmatic guardrails complement LLM outputs.
When context evidence indicates that necessary information is missing, that evidence is carried forward as clarification evidence rather than exposed as a separate user-visible flow.
Prior-turn intent-routing metadata is fed back as assistant content so relevant context accumulates across turns; scene changes reset accumulated state to avoid cross-scene contamination.

\subsection{Dual-Path Evidence Construction with Rule-Derived Evidence}
\label{sec:rule-fusion}
In the parallel evidence path, general FAQ retrieval and rule-derived evidence construction are launched together. The latter is a scenario-conditioned evidence-construction path: it reads context fields via the Model Context Protocol (MCP)~\cite{modelcontextprotocol2024} and feeds them to the rule engine:
\begin{equation}
\label{eq:parallel}
  \begin{aligned}
  \bigl(\mathcal{E}^{\mathrm{faq}}_{t}, \mathcal{E}^{\mathrm{rule}}_{t}\bigr)
  = \operatorname{ParallelGather}\bigl(&
    \operatorname{HybridRetrieve}(\tilde{q}_{t}),\\
  & \operatorname{RuleEvidence}(\operatorname{MCP}(s_{t},\eta_{t}))
  \bigr),
  \end{aligned}
\end{equation}
\texttt{ParallelGather} returns general FAQ evidence and rule-derived evidence under the same candidate-evidence contract.
In our deployment, MCP is therefore not an answer source or an autonomous decision path: it is the context-access layer that supplies permissioned fields to the rule engine.

\paragraph{Symbolic rule matching.}
Let $\phi_t$ denote the normalized context-field record returned through the MCP context layer at turn $t$. Each policy rule $r$ defines a predicate over $\phi_t$:
\begin{equation}
\label{eq:rule-match}
  P_r(\phi_t)
  = \bigwedge_{k}
    \operatorname{EvalSub}_{r,k}\bigl(\phi_t\bigr),
\end{equation}
where \texttt{EvalSub} implements field resolvers and operators. The matched rule set is
\begin{equation}
\label{eq:rule-set}
  \mathcal{R}_{t} = \{ r : P_r(\phi_t)=1 \}.
\end{equation}
Each $r \in \mathcal{R}_{t}$ is converted to a KB-shaped rule-derived evidence record and prepended as auxiliary scenario evidence:
\begin{equation}
\label{eq:candidate-merge}
  \mathcal{C}_{t}
  = \operatorname{RulesToEvidence}(\mathcal{R}_{t})
    \,\|\, \mathcal{E}^{\mathrm{faq}}_{t},
\end{equation}
where $\|$ denotes list concatenation with rule-evidence rows first. Rule matches are exposed as scenario evidence rather than returned directly, so even $|\mathcal{R}_{t}|{=}1$ flows to $\Delta$ in~\eqref{eq:decide} rather than forcing an immediate answer. In other words, MCP-accessed context fields are translated into rule-derived evidence records, then judged together with FAQ candidates by the final issue/action selection stage. A small class of state-qualified guardrail outcomes can bypass final FAQ election when evidence construction cannot safely resolve the turn; these outcomes are governed by the deterministic state policy in Section~\ref{sec:decision}.

\paragraph{Unified candidate contract.}
The novelty is not the existence of multiple retrievers, but the evidence contract that all candidates must satisfy before the decision LLM can use them.
Each item in $\mathcal{C}$ carries: (i) a stable FAQ issue ID when it is a FAQ candidate, or a rule-evidence identifier when it is scenario evidence, (ii) a category such as FAQ or rule evidence, (iii) provenance fields identifying the retrieval channel, rule ID, or scenario, (iv) rank signals such as RRF score, reranker score, and candidate position when available, and (v) an allowed action surface (\labeltext{Answer}, \labeltext{Clarify}, \labeltext{Human Handoff}) rather than an unconstrained response.
This contract has three practical effects.
First, it makes heterogeneous evidence comparable without forcing backend JSON into a free-form prompt.
Second, it supports partial evidence: when only a subset of retrieval, rerank, or MCP evidence is available, the remaining candidates still satisfy the same interface.
Third, it makes the decision trace auditable because the final issue ID and action can be tied back to an explicit candidate record.
The decision prompt therefore asks the LLM to resolve ambiguity among grounded candidates, not to discover hidden evidence sources or infer business policy from raw logs.

\paragraph{Unified clarification action.}
Clarification can be triggered by multiple evidence gaps, including missing scenario-specific context fields or ambiguity among FAQ candidates. These evidence gaps are carried into the final decision prompt, which decides whether to answer with grounded evidence, ask a necessary \labeltext{Clarify}, or follow the human-handoff policy. Counters bound repeated clarification turns, so the model can ask, while workflow state decides when asking stops.

\subsection{Hybrid Retrieval and Ranking}
\label{sec:hybrid}
FAQ retrieval executes three channels, with BM25 and embedding fetches started in parallel:
\begin{itemize}[leftmargin=1.2em]
  \item BM25 over the issue title and issue description fields (fetch 150),
  \item KNN over issue-title embeddings (fetch 10),
  \item KNN over issue-description embeddings (fetch 10).
\end{itemize}
Available hits feed RRF fusion.
For channel $c \in \{\mathrm{BM25}, \mathrm{text}, \mathrm{desc}\}$, let $L_c = (d_{c,1}, d_{c,2}, \ldots)$ be the ranked hit list returned by OpenSearch.
BM25 uses weighted multi-field matching over the issue title and issue description fields, with the issue title field boosted by $2{\times}$; KNN channels query $\mathbf{e}(\tilde{q})$ against issue-title and issue-description embeddings indexed with Qwen3-Embedding-0.6B.

\begin{figure}[t]
  \centering
  \includegraphics[width=\columnwidth]{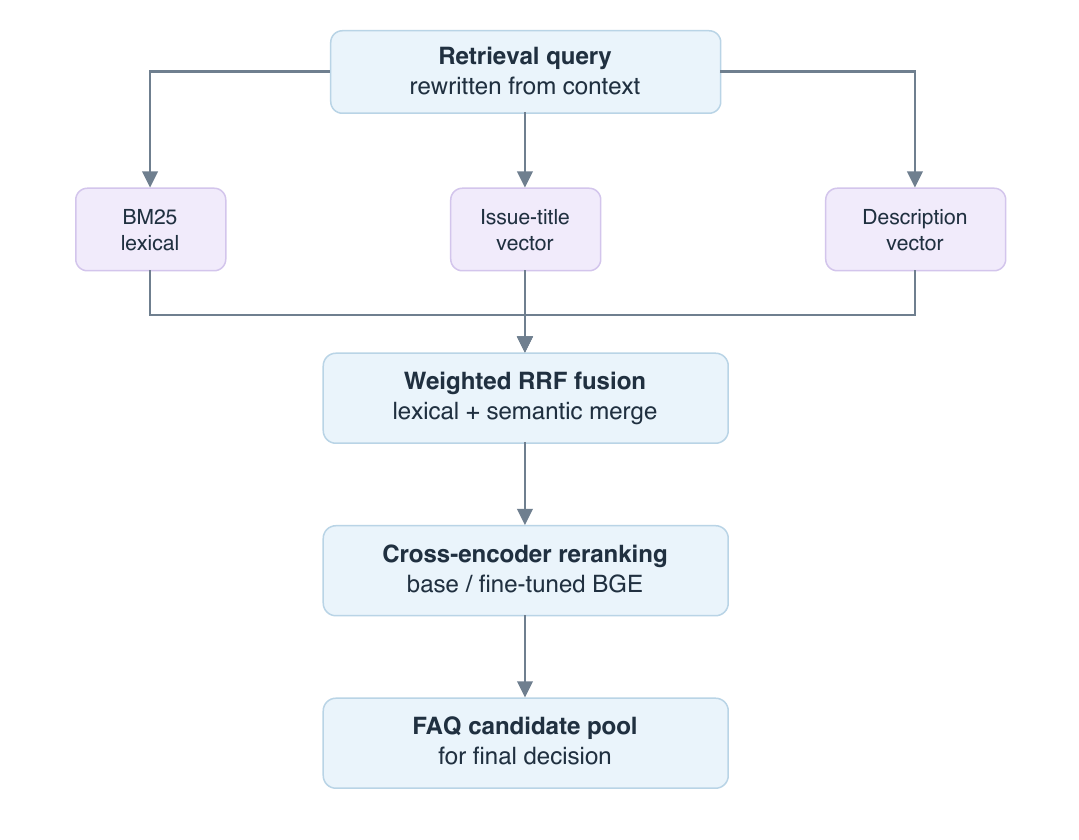}
  \caption{Three-channel hybrid FAQ recall with weighted RRF and optional reranking.}
  \label{fig:retrieval_pipeline}
\end{figure}

Figure~\ref{fig:retrieval_pipeline} expands the RAG evidence branch in Figure~\ref{fig:workflow_pipeline}: the rewritten query fans out to lexical and semantic channels, RRF merges their evidence, and the cross-encoder reranker produces the FAQ candidate pool consumed by the final issue/action selection stage.

After RRF fusion, each document $d$ (keyed by stable FAQ ID) receives a weighted reciprocal rank fusion score with per-channel deduplication (a channel contributes at most once per ID):
\begin{equation}
\label{eq:rrf}
  S_{\mathrm{RRF}}(d) = \sum_{c \,:\, d \in L_c} \frac{w_c}{K + \operatorname{rank}_c(d)},
\end{equation}
with a fixed production configuration $K{=}40$, $w_{\mathrm{BM25}}{=}1.05$, $w_{\mathrm{text}}{=}0.85$, and $w_{\mathrm{desc}}{=}1.0$~\cite{cormack2009rrf,chen2022hybrid,bruch2023fusion}. Candidates are sorted by $S_{\mathrm{RRF}}$ and truncated to the pre-rerank pool size $N_{\mathrm{RRF}}{=}50$.

\paragraph{Cross-encoder reranking.}
Let $\pi_{\mathrm{RRF}}$ be the RRF ordering. When reranking is enabled, the reranker scores pairs $(\tilde{q}, \operatorname{doc}(d))$ with cross-encoder score function $s_{\mathrm{CE}}$ and induces
\begin{equation}
\label{eq:rerank}
  \pi_{\text{out}} =
    \operatorname*{argsort}_{d \in \pi_{\mathrm{RRF}}}\,
    s_{\mathrm{CE}}\bigl(\tilde{q}, \operatorname{doc}(d)\bigr).
\end{equation}
Here $\operatorname*{argsort}$ orders candidates by descending cross-encoder score. When reranking is disabled, $\pi_{\text{out}}=\pi_{\mathrm{RRF}}$.
Final FAQ hits are $\mathcal{C}_{\text{faq}} = \pi_{\text{out}}[:k]$.
Offline evaluation (E2, Section~\ref{sec:exp-rerank}) shows that the base BGE reranker~\cite{xiao2024bge,reimers2019sentencebert} has weak top-1 ranking on business FAQ candidates, whereas the distilled fine-tuned reranker improves held-out business Hit@1 and MRR without changing the serving architecture---motivating a distilled fine-tuned checkpoint in production.

\paragraph{Related-question presentation.}
The ranked FAQ pool is also reused for user-facing related question suggestions. After the final answer is generated, selected neighboring candidates from the same retrieved and reranked evidence pool can be rendered below the response as related questions. Internal issue IDs, issue descriptions, replies, rule evidence, and retrieval scores are not exposed to users. This reuse keeps the suggestion layer aligned with the same evidence used by the customer-service agent while offering adjacent self-service paths when the user's problem is broader than the selected answer.

\FloatBarrier
\subsection{State-Constrained LLM Decision Layer}
\label{sec:decision}
The decision stage implements~\eqref{eq:decide}: $(\hat{\iota}, \hat{a}) = \Delta(\mathcal{C}, q_t, H)$ where $\mathcal{C}$ follows~\eqref{eq:candidate-merge}; when no rule-derived candidate is available, $\mathcal{C}$ simply contains the general FAQ candidates. E1 (Section~\ref{sec:exp-results}) evaluates this mapping in isolation: with the base BGE reranker, Hit@1 rises from $\approx$57\% to $\approx$81\% when the LLM reads retrieved FAQ candidates and rule-evidence blocks---evidence that $\Delta$ should not be replaced by $\pi_{\text{out}}[1]$ alone.

\paragraph{Programmatic state and human handoff.}
Turn metadata $\mathcal{M}_t$ augments the LLM with persistent counters:
\begin{equation}
\label{eq:state}
  \mathcal{M}_{t+1}
  = \mathcal{F}\bigl(
      \mathcal{M}_{t},\;
      a_{\text{route},t},\;
      \hat{a},\;
      s_{t}
    \bigr),
\end{equation}
where $\mathcal{F}$ updates clarification and handoff-request state using the current scene state $s_t$ and merges accumulated context; scene changes reset state to avoid cross-scene leakage. Human handoff is therefore a state-qualified outcome rather than a single-turn keyword decision: routing signals propose the need for escalation, while deterministic state policy decides whether escalation is appropriate. These policy thresholds are serving configuration, not experimental claims.
The session-analysis stage enforces $|H'| \leq H_{\max}$ turns and $|\text{chars}(H')| \leq L_{\max}$ before any LLM call.
On \labeltext{Answer}, the system resolves reply text from the elected issue ID, optionally wraps it with an LLM summary and collapsible KB source, and may persist QA pairs when the audit configuration enables it.

\section{System Deployment}
\label{sec:deployment}
The workflow is deployed as an online API integrated with platform risk controls and observability infrastructure. The deployment is not a prompt wrapper around an LLM: every production turn passes through policy-guided checks for MCP/rule-evidence access, context sufficiency, evidence construction, action selection, and logging. We emphasize deployment lessons that complement the offline FAQ metrics:

\paragraph{Channel isolation.}
The MCP context branch and OpenSearch retrieval run as isolated evidence branches. When only one branch contributes evidence, the decision stage still receives typed candidates rather than a malformed prompt.

\paragraph{Configuration-controlled serving.}
Feature flags control OpenSearch recall, KB-service routing, reranker enablement, and answer formatting modes. The production reranker component can use a fine-tuned BGE checkpoint rather than the base BGE checkpoint~\cite{xiao2024bge}, based on held-out business ranking checks, public reranking benchmarks, and B-Set replay.

\paragraph{Auditability.}
Each request emits a chain log tracing intent-routing JSON, MCP status, rule hit list, FAQ recall scores, and final decision rationale---supporting post-hoc review in regulated customer-service settings.

\paragraph{Trace-driven improvement loop.}
Workflow traces localize failures to intent routing, clarification, rule-evidence coverage, rule triggers, recall, reranking, final candidate selection, or action policy instead of treating wrong answers as generic LLM mistakes. This attribution determines the update target: KB content and query rewriting for recall misses, hard negatives or model choice for reranker mismatches, rule edits for policy mismatches, clarification-prompt or context-handling edits for clarification failures, and action-policy examples for handoff mistakes. E2 is one concrete instance: logs exposed reranker head quality as the dominant bottleneck, leading to in-domain reranker adaptation rather than backbone scaling.

The loop is intentionally conservative. Production traces do not directly rewrite rules, prompts, or KB entries at serving time; they produce review queues and replay sets. This is closer to a customer-support data flywheel~\cite{zhao2025agentloop} than to autonomous self-modification, preserving auditability and rollback.

\paragraph{Support-agent-assisted evidence annotation.}
After deployment, unresolved conversations are routed to human support. We embed evidence annotation into this support workflow rather than treating it as a separate offline labeling task. When a handoff occurs, the support workspace shows the current user query, necessary conversation context, the agent's attempted response, and an evidence panel containing both retrieved FAQ candidates and rule-derived evidence. FAQ candidates are expanded from RRF@50 and displayed with their issue ID, issue description, and reply, so support agents can judge whether a candidate directly resolves the user's problem.

The annotation output is structured by update target. If a correct FAQ is present but ranked poorly, the selected issue becomes the positive label and the top-ranked remaining candidates are used as hard negatives, producing reranker supervision with one positive and seven hard negatives. If rule-derived evidence is misleading, the trace is routed to evidence-condition supervision for updating evidence conditions or mappings. If the evidence is available but the final LLM decision is wrong, the trace becomes decision-level supervision: the corrected issue ID or \labeltext{Clarify} decision can serve as the chosen response, while the wrong issue ID or unnecessary clarification becomes the rejected response for future preference-tuning studies. Human handoff is treated more conservatively because it is primarily governed by policy and conversation state. If no evidence covers the query, the trace is marked as a recall miss or knowledge gap and is routed to retrieval improvement or FAQ maintenance. All updates are validated through replay before integration.

\begin{figure*}[t]
  \centering
  \includegraphics[width=0.96\textwidth]{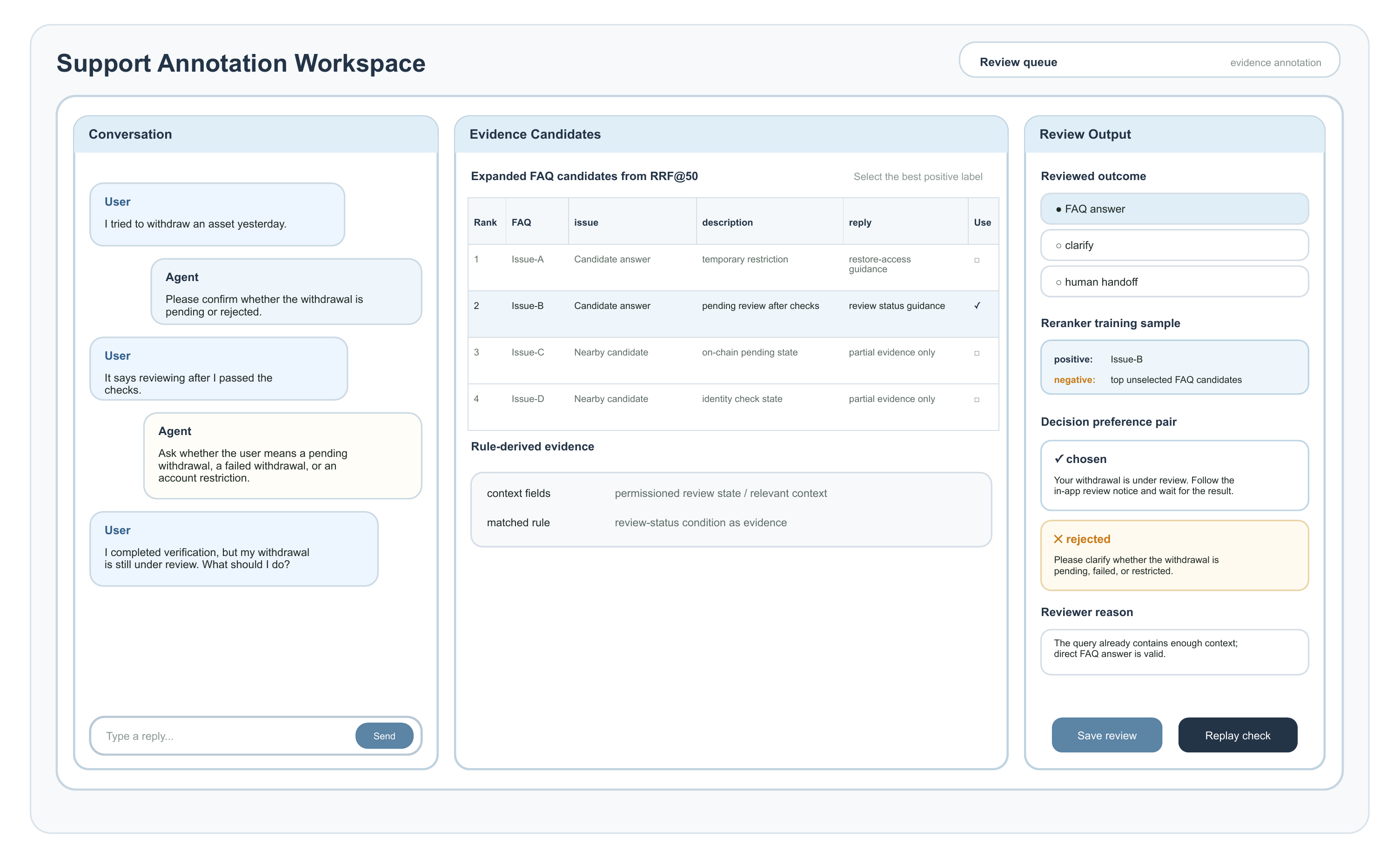}
  \caption{Anonymized support-agent workspace with evidence annotation. Support agents review the masked user query, conversation context, the agent's attempted response, RRF@50 FAQ candidates, and rule-derived evidence. The same workspace produces reranker supervision, evidence-condition fixes, decision-level preference pairs, or recall/FAQ update signals, all validated by replay before integration.}
  \label{fig:support_annotation}
\end{figure*}

\paragraph{Safety gates.}
Input and output guards, URL risk filtering, and image moderation run at API boundaries independent of the retrieval pipeline.

\section{Experiments}
\label{sec:experiments}

\subsection{Evaluation Scope and Claims}
\label{sec:eval-scope}

This paper is an \textbf{Applied Data Science} account of a deployed Evidence-Grounded Customer-Service Agent Workflow. The method sections describe the full DAG, while four evaluation blocks isolate the components that can be labeled reliably: pipeline diagnosis (E1), reranker optimization (E2), B-Set session validation (E3), and final-decision preference optimization (E4). Table~\ref{tab:canonical-metrics} maps each block to its protocol and claim. Clarification precision, rule-trigger precision, MCP/rule-evidence coverage, and full action accuracy are not yet comprehensively labeled in the offline set.

\paragraph{Human-AI evaluation workflow.}
The evaluation process follows a human-AI collaborative workflow: automated replay jobs generate candidate lists, rerank traces, and preliminary error attribution, while human reviewers define valid labels, inspect boundary cases, and approve the final B-Set review rubric. Thus the paper reports not only model scores, but also the trace evidence that explains which subsystem should be changed next.

\subsection{Evaluation Blocks and Headline Metrics}
\label{sec:canonical-metrics}

Table~\ref{tab:canonical-metrics} separates \textbf{which evidence supports which claim}, so diagnostic ablations, public benchmark checks, B-Set review, and final-decision LLM post-training are not collapsed into a single score. Detailed Hit@$K$ curves are reported in later tables; this table is a roadmap for the evaluation blocks.

\begin{table*}[t]
  \caption{Canonical evidence map for the stepwise evaluation. Detailed Hit@$K$ tables are reported in the corresponding E1/E2/E3 subsections.}
  \label{tab:canonical-metrics}
  \centering
  \footnotesize
  \setlength{\tabcolsep}{3pt}
  \adjustbox{width=0.95\textwidth,center}{%
  \begin{tabular}{@{}>{\raggedright\arraybackslash}p{0.14\textwidth}>{\raggedright\arraybackslash}p{0.21\textwidth}>{\raggedright\arraybackslash}p{0.20\textwidth}>{\raggedright\arraybackslash}p{0.19\textwidth}>{\raggedright\arraybackslash}p{0.14\textwidth}@{}}
    \toprule
    \textbf{Block} & \textbf{Question} & \textbf{Dataset / protocol} & \textbf{Key result} & \textbf{Use in paper} \\
    \midrule
    E1 Pipeline diagnosis & Is the baseline bottleneck retrieval, reranking, or the generator? & T-Set non-test diagnostic subset ($N{=}309$) and module ablation ($N{=}544$). & Recall top-50 is high; base reranker top-1 accuracy is weak; diagnostic issue-election differs by $<$1pp across Qwen/GPT-4o. & Locate evidence loss before changing models. \\
    E2 Reranker optimization & Does distilled reranker tuning improve evidence ordering without obvious benchmark regression? & T-Set 492/52/37 split and selected six-task C-MTEB proxy. & Test37 Hit@1 improves 56.76$\to$75.68; C-MTEB Avg. reaches 66.58. & Select the distilled reranker and monitor forgetting risk. \\
    E3 B-Set validation & Does better evidence improve user-facing sessions? & B-Set, 200 multi-turn sessions. & Base reranker 86.5\%; distilled reranker 88.5\% KB-grounded session accuracy. & Headline session-level validation. \\
    E4 Decision-stage DPO & Can final-decision post-training improve issue/clarification election under a fixed evidence path? & B-Set with intent routing, retrieval/RRF, candidate format, and review rubric fixed. & Isolated base-reranker + DPO reaches 90.5\%; best combined distilled-reranker + DPO reaches 92.5\%. & Isolate final-decision LLM optimization; report best system configuration. \\
    \bottomrule
  \end{tabular}}
\end{table*}

\subsection{Shared Dataset and Protocol}
\label{sec:exp-setup}

\textbf{Data roles.} Table~\ref{tab:data-roles} summarizes the data used in the evaluation. For offline development, we curate a historical labeled-trace sample containing 605 anonymized dialogue sessions and 1,427 turns. After cleaning, deduplication, and FAQ-issue filtering, its T-Set FAQ subset contains 581 turn-level FAQ queries: 492/52/37 are used for reranker train/development/held-out test, the 544 non-test queries support module-level ablations, and a 309-query non-test diagnostic subset supports early controlled pipeline diagnosis. The 309-query diagnostic subset is used only for early pipeline diagnosis and is not treated as an independent held-out evaluation set. B-Set is a sealed multi-turn holdout for KB-grounded session review.

\begin{table*}[t]
  \caption{Dataset roles used throughout the paper. T-Set FAQ rows are turn-level issue queries; B-Set is session-level validation and also contains action labels.}
  \label{tab:data-roles}
  \centering
  \footnotesize
  \setlength{\tabcolsep}{4pt}
  \adjustbox{width=0.95\textwidth,center}{%
  \begin{tabular}{@{}lrrp{0.17\textwidth}p{0.38\textwidth}@{}}
    \toprule
    \textbf{Set} & \textbf{Sessions / cases} & \textbf{Turns / queries} & \textbf{FAQ / issue subset} & \textbf{Use / caveat} \\
    \midrule
    T-Set curated sample & 605 & 1,427 & 581 & Curated historical labeled traces; the FAQ subset is used for reranker work. \\
    T-Set non-test diagnostic subset & --- & 309 & 309 & Early FAQ-issue diagnosis; non-test development subset. \\
    T-Set non-test ablation subset & --- & 544 & 544 & Module-level recall/ranking ablations; excludes the held-out Test37 split. \\
    T-train & --- & 492 & 492 & Reranker training; issue-only, not an action-evaluation set. \\
    T-dev & --- & 52 & 52 & Reranker development and checkpoint selection. \\
    T-test & --- & 37 & 37 & Held-out business FAQ reranker test. \\
    B-Set & 200 & 632 & 409 ranking-evaluable issue turns & Sealed session-level validation; evaluates FAQ, clarification, and handoff behavior. \\
    \bottomrule
  \end{tabular}}
\end{table*}

\textbf{Query construction.} User utterances are machine-translated to English. For \emph{retrieval}, all user turns in a session are concatenated into a single query string (chronological order) and passed to three-channel hybrid recall, which returns up to 50 FAQ candidates. For \emph{decision}, the decision-stage LLM receives only the \emph{last} user turn as the active query, together with the recalled candidate list---mirroring production behavior where the latest utterance disambiguates intent while prior turns inform recall breadth.

\textbf{Decision models in E1.} The diagnostic comparison uses Qwen3.5-27B and Azure GPT-4o under the same pipeline setting. Both emit structured JSON containing the selected issue ID, action, and rationale. Reranking uses the off-the-shelf BGE reranker checkpoint and reorders the recall top-50 before LLM candidate selection over \textbf{top-20} candidates. A top-10 diagnostic run is used only to understand candidate-window sensitivity.

\textbf{Reranker protocol in E2.} Hybrid recall returns top-50 candidates, which are rescored by a cross-encoder reranker over FAQ issue, description, and reply text. We compare the base BGE reranker against supervised and teacher-student distilled fine-tuned checkpoints trained on query--FAQ pairs: each training query has one positive FAQ labeled by its gold issue ID and negatives sampled from the recall pool. Generalization is checked on the cleaned held-out business split, public C-MTEB tasks, and B-Set review. Additional cleaned-data variants are kept in the experiment archive and reported only when they clarify forgetting risk.

\textbf{Metric.} We report \textbf{Hit@$K$}: the fraction of cases whose gold issue ID appears within the top-$K$ ranked candidates at a given pipeline stage. Hit@1 at the LLM stage is \emph{end-to-end issue ID accuracy}---the model's elected issue ID matches the label. At recall, Hit@50 indicates whether the gold issue appears anywhere in the retrieved pool (recall ceiling before reranking/LLM).

\paragraph{B-Set profile.}
Because the deployed system predicts both FAQ issue IDs and dialogue actions, the B-Set is not a pure single-label FAQ-matching benchmark. Table~\ref{tab:dataset-label-dist} reports the dataset-level profile used for end-to-end validation. Ranking denominators are reported separately in Table~\ref{tab:bset-issue-ranking}, because ranking metrics require turns with FAQ issue targets and available retrieval/rerank evidence.

\begin{table}[t]
  \caption{B-Set dataset profile. B-Set is used for KB-grounded session review; issue-ranking denominators are reported separately because ranking metrics apply only to turns with FAQ issue targets and retrieval/rerank evidence.}
  \label{tab:dataset-label-dist}
  \centering
  \footnotesize
  \setlength{\tabcolsep}{4pt}
  \begin{tabular*}{\columnwidth}{@{\extracolsep{\fill}}lrrr>{\raggedright\arraybackslash}p{0.48\columnwidth}@{}}
    \toprule
    \textbf{Set} & \textbf{Sessions} & \textbf{Turns} & \textbf{Avg. turns} & \textbf{Validation role} \\
    \midrule
    B-Set & 200 & 632 & 3.16 & Sealed multi-turn holdout with FAQ, clarification, and handoff labels; session accuracy requires all turns in a session to pass. \\
    \bottomrule
  \end{tabular*}
\end{table}

\paragraph{Early fixed-evidence prompt replay.}
Table~\ref{tab:prompt-loop} reports the early loop used to settle the final-decision prompt before the main B-Set evaluation; it is an auxiliary development check, not a B-Set headline result. The scoped loop follows the same engineering pattern used elsewhere in our replay workflow: define the fixed-evidence decision rubric, set the goal of improving final-decision accuracy without regressions, replay the base prompt, attribute old-fail/new-pass and old-pass/new-fail cases, apply small human-approved prompt patches across rounds, rerun the same subset, and retain the best approved prompt. The replay subset is sampled from the curated historical T-Set data and freezes query context, rewritten query, top-10 FAQ candidates, expanded FAQ text, early-exit decisions, and the grader; only the final-decision prompt changes.

\begin{table}[t]
  \caption{Early fixed-evidence prompt replay on a T-Set development subset ($200$ sessions / $578$ turns). Only the final-decision prompt changes; results are auxiliary rather than B-Set headline metrics.}
  \label{tab:prompt-loop}
  \centering
  \footnotesize
  \setlength{\tabcolsep}{3pt}
  \begin{tabular*}{\columnwidth}{@{\extracolsep{\fill}}lrr@{}}
    \toprule
    \textbf{Round / applied patch} & \textbf{Turn Acc. (\%)} & \textbf{Session Acc. (\%)} \\
    \midrule
    round0 / base prompt & 89.8 & 79.0 \\
    round1 / direct matching & 92.4 & 83.0 \\
    round2 / context and parent FAQ & 92.4 & \textbf{83.5} \\
    round3 / unsupported-claim guardrail & \textbf{92.6} & \textbf{83.5} \\
    \bottomrule
  \end{tabular*}
\end{table}

\subsection{E1: Pipeline Diagnosis}
\label{sec:exp-results}

\noindent\textit{Purpose.} E1 answers the first diagnostic question: \emph{is low accuracy caused mainly by the decision LLM, or by the evidence it receives?} We combine three controlled views: an early T-Set non-test diagnostic subset, T-Set retrieval/reranking ablation, and a decision-model comparison. The goal is to localize the bottleneck before changing the reranker, prompt, or generator.

Table~\ref{tab:retrieval-ablation} checks the retrieval/ranking stack on the 544 non-test T-Set FAQ queries, excluding the held-out Test37 split. It shows why the production path uses hybrid recall, RRF fusion, and a cross-encoder reranker. Table~\ref{tab:exp-hitk} then reports Hit@$K$ curves on the 309-query T-Set non-test diagnostic subset and compares issue Hit@1 across decision backbones: under this diagnostic pipeline, the $<$1pp gap between Qwen3.5-27B and GPT-4o suggests that raising end-to-end accuracy on this benchmark requires better evidence ordering, not a larger generator alone.

\begin{table}[t]
  \caption{T-Set module-level recall and ranking ablation ($N{=}544$ non-test turn-level FAQ queries labeled with gold FAQ issue IDs; Test37 excluded). Dense recall uses issue-title and issue-description vectors; RRF fuses BM25, issue-title-vector, and issue-description-vector channels before cross-encoder reranking.}
  \label{tab:retrieval-ablation}
  \centering
  \scriptsize
  \setlength{\tabcolsep}{2.4pt}
  \begin{tabular*}{\columnwidth}{@{\extracolsep{\fill}}lrrrrrr@{}}
    \toprule
    \textbf{Variant} & \textbf{@1} & \textbf{@5} & \textbf{@10} & \textbf{@20} & \textbf{@50} & \textbf{MRR} \\
    \midrule
    BM25 lexical recall & 41.9 & 70.0 & 82.2 & 88.4 & 95.0 & 0.5434 \\
    Dense recall & 30.3 & 61.9 & 76.5 & 88.8 & 88.8 & 0.4492 \\
    RRF fusion & 41.2 & 74.1 & 86.8 & 94.1 & \textbf{99.3} & 0.5532 \\
    Base BGE reranker & \textbf{59.0} & \textbf{89.2} & \textbf{95.2} & \textbf{97.6} & \textbf{99.3} & \textbf{0.7181} \\
    \bottomrule
  \end{tabular*}
\end{table}

\paragraph{RRF and reranker configuration.}
We set RRF to $K{=}40$ with channel weights $(w_{\mathrm{BM25}}, w_{\mathrm{text}}, w_{\mathrm{desc}})=(1.05, 0.85, 1.00)$ and fetch sizes $150/10/10$ for BM25, issue-title vector, and issue-description vector recall. Query and indexed FAQ vectors use Qwen3-Embedding-0.6B. This setting reaches the highest T-Set top-50 coverage among the representative configurations we tested; BM25-only, dense-only, smaller candidate budgets, and over-expanded dense fetch all reduce top-50 coverage or add cost without improving the downstream candidate pool. For the cross-encoder base, we select BGE-reranker-large: with full FAQ payloads and a shared 500-token budget, it reaches 95.2\% Hit@10 and 97.6\% Hit@20 on T-Set, while improving MRR from the RRF-only 0.5532 to 0.7181. BGE-reranker-v2-m3 improves Hit@20 only slightly (98.3\%) and shows higher latency risk when allowed to use its longer native context window.

\begin{table}[t]
  \caption{Issue ID Hit@$K$ by pipeline stage on the $N{=}309$ T-Set non-test diagnostic subset. Recall: concatenated multi-turn user queries. LLM: last user turn only.}
  \label{tab:exp-hitk}
  \centering
  \scriptsize
  \setlength{\tabcolsep}{1.5pt}
  \renewcommand{\arraystretch}{1.04}
  \begin{tabular*}{\columnwidth}{@{\extracolsep{\fill}}llrrrrrr@{}}
    \toprule
    Setting & Stage & @1 & @5 & @10 & @15 & @20 & @50 \\
    \midrule
    Base BGE/Qwen & Hybrid recall & 42.07 & 74.43 & 84.47 & 88.35 & 90.29 & 96.76 \\
    Base BGE/Qwen & Rerank (top-50) & 56.31 & 84.14 & 91.59 & 94.50 & 95.15 & --- \\
    Base BGE/Qwen & LLM (Qwen3.5-27B) & \textbf{81.23} & --- & --- & --- & --- & --- \\
    \midrule
    Base BGE/GPT-4o & Hybrid recall & 42.53 & 74.03 & 83.77 & 87.66 & 89.94 & 96.75 \\
    Base BGE/GPT-4o & Rerank (top-50) & 56.82 & 83.77 & 91.56 & 94.48 & 95.13 & --- \\
    Base BGE/GPT-4o & LLM (Azure GPT-4o) & \textbf{81.88} & --- & --- & --- & --- & --- \\
    \bottomrule
  \end{tabular*}
\end{table}

\paragraph{Stage-wise interpretation.}
Table~\ref{tab:exp-hitk} decomposes where error remains under a \emph{base BGE} reranker:
\begin{itemize}[leftmargin=1.2em]
  \item \textbf{Recall ($\approx$42\% Hit@1, $\approx$97\% Hit@50):} hybrid fusion finds the gold FAQ in the top-50 pool for most cases but rarely at rank~1---the system should not terminate at recall.
  \item \textbf{Reranking ($\approx$57\% Hit@1):} generic cross-encoder reranking helps (+15pp) yet $\approx$44\% of cases still have a wrong top-1 candidate.
  \item \textbf{Candidate election ($\approx$81\% Hit@1):} the decision-stage LLM reads candidate summaries and the latest user turn, recovering +25pp over reranking---primary offline evidence that the decision flow should not stop at retrieve-then-rank when top-1 ranking is weak.
\end{itemize}
Under the diagnostic pipeline, replacing Qwen3.5-27B with GPT-4o changes final issue-election accuracy by less than 1pp, suggesting that backbone scaling alone is not the dominant bottleneck in this setting.

\paragraph{Candidate window.}
The early T-Set non-test diagnostic subset used top-20 evidence to test whether LLM candidate selection can recover from weak top-1 ranking. A top-10 diagnostic run under the same base reranker lowers decision Hit@1 from 81.23\% to 78.96\%, showing that ranks 11--20 can compensate for a weak generic reranker. After reranker adaptation, the compact top-10 window is preferred for serving because the T-Set and Test37 results show that the gold FAQ is already usually inside the top-10 evidence window.

\subsection{E2: Teacher-Student Reranker Fine-Tuning}
\label{sec:exp-rerank}

\noindent\textit{Purpose.} E2 is the intervention suggested by the funnel diagnosis. Since decision LLMs cannot fully compensate for weak top-1 ranks when $\approx$44\% of top-1 rerank results are wrong, we ask whether a reranker fine-tuned on in-domain FAQ pairs can improve top-rank ordering on held-out business supervision---\emph{without} involving the LLM---and at what cost to public reranking benchmarks. After training, we evaluate each checkpoint on (a)~public reranking benchmarks and (b)~the cleaned 492/52/37 train/development/test business split.

\paragraph{Training-data cleaning.}
Each reranker supervision group contains one positive FAQ and seven hard negatives sampled from the recall pool. The teacher-student run starts from 506 candidate training groups and removes 14 groups before training, leaving 492 training groups with the same 52/37 development/test split. Removed groups are cases where the query and positive FAQ are visibly mismatched, the teacher margin is too low to provide a stable preference, multiple unrelated intents are forced into one positive, or the label comes from a weak human-correction path rather than a high-confidence FAQ target. We do not promote the 14 removed groups to a test set because their labels are judged unreliable; uncleaned runs are retained in the experiment archive as sensitivity controls.

\paragraph{Teacher-score distillation.}
The distilled reranker is initialized from the same BGE checkpoint as the supervised run. For each query and recalled candidate set, the teacher provides soft relevance scores; the student is optimized with the supervised positive/negative ranking objective plus a teacher-score distillation term, so that in-domain supervision improves FAQ ordering without discarding the teacher's broader ranking preferences.

\begin{table}[t]
  \caption{Internal held-out reranker test on the same 37 business FAQ groups. The saturated Hit@10 supports using top-10 as the compact FAQ evidence window; Hit@1 and MRR show whether the gold FAQ is ranked early enough to reduce downstream ambiguity.}
  \label{tab:internal-rerank-test37}
  \centering
  \footnotesize
  \setlength{\tabcolsep}{3pt}
  \begin{tabular*}{\columnwidth}{@{\extracolsep{\fill}}lrrrrr@{}}
    \toprule
    \textbf{Reranker} & \textbf{@1} & \textbf{@5} & \textbf{@10} & \textbf{@20} & \textbf{MRR} \\
    \midrule
    Base BGE & 56.76 & 86.49 & 97.30 & 97.30 & 0.6991 \\
    Supervised fine-tuned reranker & 64.86 & 86.49 & 94.59 & \textbf{100.00} & 0.7391 \\
    Distilled fine-tuned reranker & \textbf{75.68} & \textbf{91.89} & 97.30 & 97.30 & \textbf{0.8215} \\
    \bottomrule
  \end{tabular*}
\end{table}

\paragraph{Internal held-out rerank test.}
Table~\ref{tab:internal-rerank-test37} evaluates the base BGE, supervised fine-tuned, and distilled fine-tuned rerankers on the same held-out Test37 split. Although this FAQ test split is small, it is strictly separated from training and development; we use it as a controlled internal ranking check, while B-Set provides session-level validation. We use top-10 as the production-style FAQ evidence window because the base BGE and distilled fine-tuned rerankers both reach 97.30\% Hit@10, indicating that a wider window is unlikely to add much candidate coverage on this split. The remaining improvement should therefore come from \emph{top-rank ordering}: the distilled fine-tuned reranker preserves the same Hit@10 while improving Hit@1 from 56.76\% to 75.68\% and MRR from 0.6991 to 0.8215.

\paragraph{General benchmark sanity check.}
We also checked six C-MTEB reranking tasks to avoid choosing a checkpoint that only memorizes the business set. The distilled fine-tuned reranker is trained on the cleaned 492 groups; the development split is used only for model selection, and the held-out business test result is reported in Table~\ref{tab:internal-rerank-test37}. Table~\ref{tab:cmteb-six} keeps only the comparison needed for the paper: official/base scores, the deployed supervised fine-tuned reranker, and a distilled fine-tuned reranker that best preserves public ranking ability. The supervised fine-tuned reranker slightly lowers the six-task average relative to the public base model, while the distilled fine-tuned reranker recovers and exceeds the base average.

We use these public tasks as forgetting probes, not target-domain objectives. Teacher-score distillation improves the C-MTEB average from 66.09 to 66.58, suggesting that it preserves general reranking quality under this selected six-task check while retaining the business-domain top-ranking gains.

\begin{table}[t]
  \caption{Selected six-task C-MTEB reranking MAP/main\_score comparison. Older and partial checkpoints are omitted from the paper table and retained in logs.}
  \label{tab:cmteb-six}
  \centering
  \scriptsize
  \setlength{\tabcolsep}{1.5pt}
  \renewcommand{\arraystretch}{1.05}
  \begin{tabular*}{\columnwidth}{@{\extracolsep{\fill}}lrrrrrrr@{}}
    \toprule
    \textbf{Model} & \textbf{T2} & \textbf{Zh2En} & \textbf{En2Zh} & \textbf{MMarco} & \textbf{CMed1} & \textbf{CMed2} & \textbf{Avg.} \\
    \midrule
    Official BGE reranker & 67.60 & 64.03 & 61.44 & 37.16 & 82.15 & 84.18 & 66.09 \\
    Supervised fine-tuned reranker & 67.55 & 64.70 & 62.93 & 33.94 & 81.11 & 83.51 & 65.62 \\
    Distilled fine-tuned reranker & \textbf{67.55} & \textbf{64.89} & \textbf{63.37} & \textbf{38.27} & \textbf{81.80} & \textbf{83.59} & \textbf{66.58} \\
    \bottomrule
  \end{tabular*}
\end{table}

\paragraph{Implications.}
Teacher-score distillation reduces observed forgetting risk on the selected public reranking proxy: after cleaning the training supervision from 506 to 492 groups, the distilled fine-tuned reranker improves the six-task C-MTEB average from 66.09 to 66.58 while keeping the business test-set Hit@10 at 97.3\%. We therefore treat distillation as the preferred checkpoint recipe, while using B-Set review to decide which checkpoint is best for end-to-end customer-service quality.

\subsection{E3: End-to-End B-Set Evaluation}
\label{sec:exp-bvolume}

\noindent\textit{Purpose.} E3 validates whether reranker and evidence-quality changes improve user-facing session quality. Ranking metrics alone are insufficient for production acceptance because a customer-service agent may need to answer with a FAQ, ask a necessary clarification, or follow an action policy depending on the dialogue context. We therefore use a \textbf{B-Set} holdout of 200 multi-turn sessions that was not used for reranker training, and report both issue-turn ranking quality and KB-grounded session-level review.

\paragraph{Review protocol.}
The B-Set gold labels are produced by a KB-grounded review process over current user query, preceding user turns, maintained FAQ issue title, issue description, reply text, and action policy. A turn is correct if the final output selects a directly applicable FAQ, asks a necessary clarification when information is insufficient, or follows the appropriate dialogue action. A clarification is rejected when the KB already provides a precise answer; an FAQ answer is rejected when it is off-topic, over-specific, or ignores required session context. Session accuracy requires every turn in the session to pass under the fixed review rubric.

\begin{table}[t]
  \caption{B-Set multi-turn evaluation. The headline metric is KB-grounded session accuracy on $N{=}200$ sessions; the same review rubric is applied to all systems. Values are point estimates, and we do not claim statistical significance. The base-reranker + DPO row isolates final-decision post-training, while the distilled-reranker + DPO row reports the best combined system configuration.}
  \label{tab:bvolume-e2e}
  \centering
  \footnotesize
  \setlength{\tabcolsep}{2pt}
  \begin{tabular*}{\columnwidth}{@{\extracolsep{\fill}}>{\raggedright\arraybackslash}p{0.43\columnwidth}>{\raggedright\arraybackslash}p{0.24\columnwidth}>{\raggedleft\arraybackslash}p{0.20\columnwidth}@{}}
    \toprule
    \textbf{System} & \textbf{Reranker} & \textbf{KB-grounded session acc.} \\
    \midrule
    Base reranker + base decision LLM & Base BGE reranker & 86.5 \\
    Distilled FT reranker + base decision LLM & Distilled FT reranker & 88.5 \\
    Base reranker + DPO final-decision LLM & Base BGE reranker & 90.5 \\
    Distilled FT reranker + DPO final-decision LLM & Distilled FT reranker & 92.5 \\
    \bottomrule
  \end{tabular*}
\end{table}

\begin{figure}[t]
  \centering
  \includegraphics[width=\columnwidth]{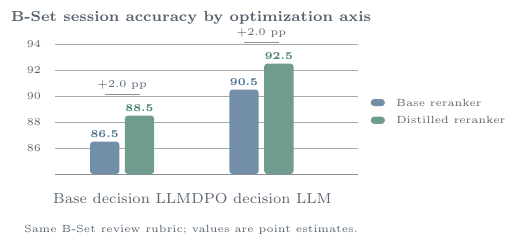}
  \caption{B-Set session accuracy under a two-axis optimization design. The grouped bars separate the reranker axis from the final-decision LLM axis under the same KB-grounded session review rubric over 200 sessions; issue-turn ranking metrics are reported separately in Table~\ref{tab:bset-issue-ranking}.}
  \label{fig:bset-session-accuracy}
\end{figure}

\paragraph{Issue-turn ranking quality.}
Table~\ref{tab:bset-issue-ranking} isolates the ranking-evaluable FAQ issue turns inside B-Set, namely turns with gold FAQ issue targets and available retrieval/rerank evidence. Base and distilled configurations use the same RRF candidate pool, so differences after RRF are attributable to reranking.

\begin{table}[t]
  \caption{B-Set issue-turn ranking quality under the same RRF candidate pool. Values are percentages over $409$ ranking-evaluable gold issue turns.}
  \label{tab:bset-issue-ranking}
  \centering
  \scriptsize
  \setlength{\tabcolsep}{2pt}
  \adjustbox{max width=\columnwidth,center}{%
  \begin{tabular*}{\columnwidth}{@{\extracolsep{\fill}}lrrrrr@{}}
    \toprule
    \textbf{Reranker} & \textbf{RRF@50} & \textbf{RRF@10} & \textbf{Rerank@1} & \textbf{Rerank@5} & \textbf{Rerank@10} \\
    \midrule
    Base & 98.3 & 93.4 & 69.4 & 91.9 & 96.6 \\
    Distilled FT & 98.3 & 93.4 & 79.2 & 95.4 & 97.8 \\
    \bottomrule
  \end{tabular*}}
\end{table}

\paragraph{Rerank-only contrast.}
Table~\ref{tab:bset-rerank-examples} gives inspected examples behind the aggregate B-Set ranking gains. On this shared candidate pool, the distilled fine-tuned reranker moves five inspected FAQ targets from outside the base reranker Top-10 into Top-10, with no reverse Top-10 regression in the inspected differences. Three gains translate into corrected final FAQ decisions; the remaining gains improve the evidence set, while the final output LLM still chooses clarification under the dialogue policy. This qualitative inspection illustrates where the ranking gain helps without turning the reranker table into an end-to-end decision metric.

\begin{table}[t]
  \caption{Inspected B-Set examples where the distilled reranker improves FAQ evidence ordering. Queries are lightly paraphrased for readability; the table is qualitative support for Table~\ref{tab:bset-issue-ranking}, not a separate headline metric.}
  \label{tab:bset-rerank-examples}
  \centering
  \scriptsize
  \setlength{\tabcolsep}{2.5pt}
  \begin{tabular*}{\columnwidth}{@{\extracolsep{\fill}}rp{0.30\columnwidth}lp{0.18\columnwidth}p{0.27\columnwidth}@{}}
    \toprule
    \textbf{Ex.} & \textbf{Representative user query} & \textbf{Base} & \textbf{Distilled} & \textbf{Downstream effect} \\
    \midrule
    1 & What is the fee for futures trading on a specific pair? & miss@10 & rank2 & Final FAQ decision becomes correct \\
    2 & My withdrawal is still under review; can I cancel it? & miss@10 & rank2 & Final FAQ decision becomes correct \\
    3 & How can I get an income-tax report? & miss@10 & rank1 & Final FAQ decision becomes correct \\
    4 & My account verification shows an unusual status. & miss@10 & rank9 & Evidence recovered; policy still clarifies \\
    5 & Can my KYC/KYB review be extended? & miss@10 & rank10 & Evidence recovered; policy still clarifies \\
    \bottomrule
  \end{tabular*}
\end{table}

\subsection{E4: Decision-Stage Preference Optimization}
\label{sec:exp-dpo}
We further evaluate whether replay-derived preference pairs can improve the final issue/action selection step. This run keeps the B-Set file, intent-routing stage, context construction, retrieval/RRF pipeline, base reranker, candidate format, review rubric, and serving policy fixed; the only changed component is the final-decision LLM used by the Evidence-Grounded Decision Module. The DPO data contains 117 high-confidence preference pairs mined from development replay failures, not from the sealed B-Set validation set: 53 \textit{wrong-issue-to-clarify}, 38 \textit{wrong-issue-to-issue}, and 26 \textit{wrong-clarify-to-issue} pairs. No B-Set turn or session is used for DPO preference pairs or checkpoint selection. Thus the preference signal is not a one-sided push toward answering with more issue IDs. It teaches three decision boundaries: reject an FAQ when it is too narrow and ask a useful clarification, avoid unnecessary generic clarification when a suitable FAQ candidate exists, and replace a mismatched FAQ with a more applicable issue. Table~\ref{tab:dpo-boundary-examples} gives anonymized examples of these boundary types.

\begin{table}[t]
  \caption{Anonymized B-Set turns where the DPO final-decision LLM corrects the baseline decision. Queries are lightly paraphrased; the table illustrates decision-boundary types rather than adding a separate headline metric.}
  \label{tab:dpo-boundary-examples}
  \centering
  \scriptsize
  \setlength{\tabcolsep}{2pt}
  \begin{tabular*}{\columnwidth}{@{\extracolsep{\fill}}>{\raggedright\arraybackslash}p{0.19\columnwidth}>{\raggedright\arraybackslash}p{0.39\columnwidth}>{\raggedright\arraybackslash}p{0.20\columnwidth}>{\raggedright\arraybackslash}p{0.16\columnwidth}@{}}
    \toprule
    \textbf{Boundary type} & \textbf{Anonymized user query} & \textbf{Baseline behavior} & \textbf{DPO behavior} \\
    \midrule
    Wrong issue $\rightarrow$ clarify & My asset was in the trading account, then I transferred it to funding, and when I checked again it disappeared. & Over-specific transfer / asset FAQ & Clarify \\
    Wrong clarify $\rightarrow$ issue & Unfreeze my withdrawal. & Generic clarification & Withdrawal restriction FAQ \\
    Wrong issue $\rightarrow$ better issue & It says the required credential uses a different format, but I see another code in my account. Should I use that? & Anti-phishing-code FAQ & Trading-password FAQ \\
    \bottomrule
  \end{tabular*}
\end{table}

Under the same B-Set review rubric, the resulting base-reranker + DPO final-decision configuration reaches 612/632 turn hits and 181/200 session hits (90.5\%). Combining the distilled reranker with the DPO final-decision LLM yields the best B-Set session accuracy of 92.5\%, but this combined row is reported as a system configuration rather than as the isolated DPO effect. This isolates a second optimization axis: after evidence quality is stable, decision-stage post-training can further calibrate the boundary between grounded FAQ answers and necessary clarifications.

\paragraph{Interpretation.}
The distilled fine-tuned reranker is therefore evaluated as an evidence-ordering change, not as a new answer generator. It improves top-rank ordering substantially on issue turns, while the final output LLM may still choose clarification when the session context is insufficient. The deployable claim is: \emph{under the same B-Set file and the same KB-grounded review rubric, the distilled fine-tuned reranker improves user-facing multi-turn session accuracy while preserving workflow guardrails against over-answering}.

\subsection{Post-Launch Operational Check}
\label{sec:post-launch-check}
We additionally examine reviewed accuracy after deployment. We draw two equally sized operational samples, each containing 1,508 dialogue sessions, from the Agent and legacy RAG-only entry paths, respectively. Each consultation is counted as one independent session, and both samples are evaluated under the same session-level operational-review rubric. The Agent workflow achieves 89.52\% reviewed accuracy, compared with 79.00\% for the legacy workflow, an observed difference of 10.52 percentage points. This result complements rather than replaces the controlled B-Set evaluation. Because the operational samples come from separate production entry paths rather than randomized assignment or matched sessions, we treat the comparison as descriptive operational evidence.

\subsection{Synthesis Across Evaluation Blocks}
\label{sec:exp-synthesis}
Table~\ref{tab:exp-synthesis} summarizes the controlled experiments as a sequence of diagnostic questions and design decisions: diagnose evidence loss (E1), improve evidence ordering (E2), validate session quality (E3), and optimize the final-decision LLM under a fixed evidence path (E4).

\begin{table}[t]
  \caption{Stepwise accuracy-improvement chain. Each row states the question asked, the evidence used, and the resulting design decision.}
  \label{tab:exp-synthesis}
  \centering
  \footnotesize
  \setlength{\tabcolsep}{3pt}
  \begin{tabular*}{\columnwidth}{@{\extracolsep{\fill}}>{\raggedright\arraybackslash}p{0.22\columnwidth}>{\raggedright\arraybackslash}p{0.34\columnwidth}>{\raggedright\arraybackslash}p{0.38\columnwidth}@{}}
    \toprule
    \textbf{Step} & \textbf{Question / evidence} & \textbf{Design decision} \\
    \midrule
    Diagnose baseline & E1: recall has high top-50 coverage, base BGE reranker is weak at rank~1, and LLM candidate selection recovers to $\approx$81\%; BM25/vector/RRF/rerank ablations identify where evidence is lost. & Keep controlled candidate selection; do not rely on reranker top-1 or generator scaling alone. \\
    Improve evidence & E2: the distilled fine-tuned reranker trained on cleaned supervision improves held-out business Hit@1/MRR, reaches 97.3\% test-set Hit@10, and improves C-MTEB Avg. to 66.58. & Move the main optimization upstream into the RAG evidence layer while monitoring forgetting. \\
    Improve decision prompt & Early fixed-evidence prompt replay: turn accuracy improves from 89.8\% to 92.6\%, and session accuracy from 79.0\% to 83.5\%. & Treat prompt edits as replay-verified development patches, not headline B-Set results. \\
    Review user outcome & E3: B-Set shows base reranker 86.5\% and distilled fine-tuned reranker 88.5\% KB-grounded session accuracy. & Report KB-grounded session review as the headline end-to-end score; use reviewed traces to update labels, KB rules, and guardrails. \\
    Optimize final decision & E4: with the base reranker fixed, DPO final-decision LLM reaches 90.5\% B-Set session accuracy. & Use replay-derived preference pairs to tune issue-vs-clarification boundaries after the evidence path is stable. \\
    Remaining gap & Comprehensive context, rule-evidence, and action accuracy remains planned. & Extend the same trace-driven method to action, context, and rule metrics after launch. \\
    \bottomrule
  \end{tabular*}
\end{table}

\section{Discussion}
The main novelty is not a new retriever, a larger chat model, or an unconstrained agent loop. It is a deployment pattern in which LLMs participate in customer-service decisions while deterministic workflow code retains authority over evidence access and policy boundaries. Hybrid RAG evidence construction keeps execution boundaries in code; evidence-grounded decision making turns retrieval results, scenario-specific rule evidence, clarification, and human handoff into auditable evidence/actions; trace-driven improvement turns traces into concrete update targets rather than a generic ``bad answer'' bucket. This explains why backbone replacement changes less than 1pp under the E1 diagnostic pipeline, while reranker adaptation is evaluated as an evidence-quality change whose user-facing value must be confirmed by B-Set review.

The trade-off is reduced autonomy. The benefit is operational control: traces reveal whether a failure came from evidence, ranking, rules, clarification, or handoff policy before an update is selected. E3 also shows why customer-service evaluation needs KB-grounded session review: valid outcomes may include answering with a FAQ, asking a necessary clarification, or following the dialogue policy.

\section{Limitations and Future Work}
Experiments focus on one financial customer-service domain and use curated internal labels rather than web-scale retrieval suites such as BEIR~\cite{thakur2021beir}. The B-Set adds a 200-session end-to-end holdout with KB-grounded human review, but it is still single-domain and manually reviewed.
Because user utterances are machine-translated to English before retrieval and evaluation, translation errors may affect recall and decision quality; multilingual end-to-end evaluation is left for future work.
Scenario-specific rule evidence, clarification, and dialogue-action labels are deployed workflow behavior but not yet comprehensively labeled in the offline set.
Controlled headline metrics follow E1/E2/E3/E4 (Table~\ref{tab:canonical-metrics}); reranker generalization is supported by the cleaned teacher-student split, public C-MTEB checks, and the B-Set review~\cite{xiao2024bge}. The separate post-launch comparison uses equally sized samples from different production entry paths under the same rubric; it is descriptive rather than a randomized causal estimate.
We do not claim autonomous ``agent'' benchmark performance~\cite{singh2025agenticrag}; the contribution is an Evidence-Grounded Customer-Service Agent Workflow~\cite{langgraph2024,chung2023instructtods,zhou2026externalization} with explicit evaluation boundaries.
Generalization beyond financial-platform customer service remains to be tested.
Future work includes broader replay evaluation, clarification/action/rule-evidence metrics, cross-domain transfer, and formal analysis of candidate fusion under noisy rule triggers.

\section*{Ethics Statement}
Customer-service automation should include explicit human-handoff safeguards, policy-compliance checks, and logging for post-hoc auditability. Sensitive user attributes should be minimized in retrieval features and logs. Automated decisions with high user impact require human-in-the-loop review paths.

\input{references.tex}

\end{document}

%% file: references.tex

%% file: arxiv_main.bbl
\begin{thebibliography}{31}


\ifx \showCODEN    \undefined \def \showCODEN     #1{\unskip}     \fi
\ifx \showISBNx    \undefined \def \showISBNx     #1{\unskip}     \fi
\ifx \showISBNxiii \undefined \def \showISBNxiii  #1{\unskip}     \fi
\ifx \showISSN     \undefined \def \showISSN      #1{\unskip}     \fi
\ifx \showLCCN     \undefined \def \showLCCN      #1{\unskip}     \fi
\ifx \shownote     \undefined \def \shownote      #1{#1}          \fi
\ifx \showarticletitle \undefined \def \showarticletitle #1{#1}   \fi
\ifx \showURL      \undefined \def \showURL       {\relax}        \fi
\providecommand\bibfield[2]{#2}
\providecommand\bibinfo[2]{#2}
\providecommand\natexlab[1]{#1}
\providecommand\showeprint[2][]{arXiv:#2}

\bibitem[{Anthropic PBC}(2024)]%
        {modelcontextprotocol2024}
\bibfield{author}{\bibinfo{person}{{Anthropic PBC}}.}
  \bibinfo{year}{2024}\natexlab{}.
\newblock \bibinfo{title}{Model Context Protocol}.
\newblock
  \bibinfo{howpublished}{\url{https://modelcontextprotocol.io/docs/getting-started/intro}}.
\newblock
\newblock
\shownote{Accessed June 2026}.


\bibitem[Asai et~al\mbox{.}(2024)]%
        {asai2024selfrag}
\bibfield{author}{\bibinfo{person}{Akari Asai}, \bibinfo{person}{Zeqiu Wu},
  \bibinfo{person}{Yizhong Wang}, \bibinfo{person}{Avirup Sil}, {and}
  \bibinfo{person}{Hannaneh Hajishirzi}.} \bibinfo{year}{2024}\natexlab{}.
\newblock \showarticletitle{Self-{RAG}: Learning to Retrieve, Generate, and
  Critique through Self-Reflection}. In \bibinfo{booktitle}{\emph{International
  Conference on Learning Representations}}.
\newblock
\urldef\tempurl%
\url{https://openreview.net/forum?id=hSyW5go0v8}
\showURL{%
\tempurl}


\bibitem[Bruch et~al\mbox{.}(2024)]%
        {bruch2023fusion}
\bibfield{author}{\bibinfo{person}{Sebastian Bruch}, \bibinfo{person}{Siyu
  Gai}, {and} \bibinfo{person}{Amir Ingber}.} \bibinfo{year}{2024}\natexlab{}.
\newblock \showarticletitle{An Analysis of Fusion Functions for Hybrid
  Retrieval}.
\newblock \bibinfo{journal}{\emph{ACM Transactions on Information Systems}}
  \bibinfo{volume}{42}, \bibinfo{number}{1}, Article \bibinfo{articleno}{20}
  (\bibinfo{year}{2024}), \bibinfo{numpages}{20:1--20:35}~pages.
\newblock
\href{https://doi.org/10.1145/3596512}{doi:\nolinkurl{10.1145/3596512}}


\bibitem[Chen et~al\mbox{.}(2022)]%
        {chen2022hybrid}
\bibfield{author}{\bibinfo{person}{Tao Chen}, \bibinfo{person}{Mingyang Zhang},
  \bibinfo{person}{Jing Lu}, \bibinfo{person}{Michael Bendersky}, {and}
  \bibinfo{person}{Marc Najork}.} \bibinfo{year}{2022}\natexlab{}.
\newblock \showarticletitle{Out-of-Domain Semantics to the Rescue! Zero-Shot
  Hybrid Retrieval Models}. In \bibinfo{booktitle}{\emph{Advances in
  Information Retrieval: 44th European Conference on IR Research, ECIR 2022}}
  \emph{(\bibinfo{series}{Lecture Notes in Computer Science},
  Vol.~\bibinfo{volume}{13185})}. \bibinfo{pages}{95--110}.
\newblock
\href{https://doi.org/10.1007/978-3-030-99736-6_7}{doi:\nolinkurl{10.1007/978-3-030-99736-6_7}}


\bibitem[Chung et~al\mbox{.}(2023)]%
        {chung2023instructtods}
\bibfield{author}{\bibinfo{person}{Willy Chung}, \bibinfo{person}{Samuel
  Cahyawijaya}, \bibinfo{person}{Bryan Wilie}, \bibinfo{person}{Holy Lovenia},
  {and} \bibinfo{person}{Pascale Fung}.} \bibinfo{year}{2023}\natexlab{}.
\newblock \showarticletitle{Instruct{TODS}: Large Language Models for
  End-to-End Task-Oriented Dialogue Systems}. In
  \bibinfo{booktitle}{\emph{Proceedings of the Second Workshop on Natural
  Language Interfaces}}. \bibinfo{pages}{1--21}.
\newblock
\href{https://doi.org/10.18653/v1/2023.nlint-1.1}{doi:\nolinkurl{10.18653/v1/2023.nlint-1.1}}


\bibitem[Cormack et~al\mbox{.}(2009)]%
        {cormack2009rrf}
\bibfield{author}{\bibinfo{person}{Gordon~V. Cormack}, \bibinfo{person}{Charles
  L.~A. Clarke}, {and} \bibinfo{person}{Stefan Buettcher}.}
  \bibinfo{year}{2009}\natexlab{}.
\newblock \showarticletitle{Reciprocal Rank Fusion Outperforms Condorcet and
  Individual Rank Learning Methods}. In \bibinfo{booktitle}{\emph{Proceedings
  of the 32nd International ACM SIGIR Conference on Research and Development in
  Information Retrieval}}. \bibinfo{pages}{758--759}.
\newblock
\href{https://doi.org/10.1145/1571941.1572114}{doi:\nolinkurl{10.1145/1571941.1572114}}


\bibitem[Gao et~al\mbox{.}(2024)]%
        {gao2024ragsurvey}
\bibfield{author}{\bibinfo{person}{Yunfan Gao}, \bibinfo{person}{Yun Xiong},
  \bibinfo{person}{Xinyu Gao}, \bibinfo{person}{Kangxiang Jia},
  \bibinfo{person}{Jinliu Pan}, \bibinfo{person}{Yuxi Bi}, \bibinfo{person}{Yi
  Dai}, \bibinfo{person}{Jiawei Sun}, \bibinfo{person}{Meng Wang}, {and}
  \bibinfo{person}{Haofen Wang}.} \bibinfo{year}{2024}\natexlab{}.
\newblock \showarticletitle{Retrieval-Augmented Generation for Large Language
  Models: A Survey}.
\newblock \bibinfo{journal}{\emph{arXiv preprint arXiv:2312.10997}}
  (\bibinfo{year}{2024}).
\newblock
\href{https://doi.org/10.48550/arXiv.2312.10997}{doi:\nolinkurl{10.48550/arXiv.2312.10997}}


\bibitem[Garcez and Lamb(2023)]%
        {garcez2023neurosymbolic}
\bibfield{author}{\bibinfo{person}{Artur~d'Avila Garcez} {and}
  \bibinfo{person}{Luis~C. Lamb}.} \bibinfo{year}{2023}\natexlab{}.
\newblock \showarticletitle{Neurosymbolic {AI}: The 3rd Wave}.
\newblock \bibinfo{journal}{\emph{Artificial Intelligence Review}}
  \bibinfo{volume}{56}, \bibinfo{number}{11} (\bibinfo{year}{2023}),
  \bibinfo{pages}{12387--12406}.
\newblock
\href{https://doi.org/10.1007/s10462-023-10448-w}{doi:\nolinkurl{10.1007/s10462-023-10448-w}}


\bibitem[Jeong et~al\mbox{.}(2024)]%
        {jeong2024adaptiverag}
\bibfield{author}{\bibinfo{person}{Soyeong Jeong}, \bibinfo{person}{Jinheon
  Baek}, \bibinfo{person}{Sukmin Cho}, \bibinfo{person}{Sung~Ju Hwang}, {and}
  \bibinfo{person}{Jong Park}.} \bibinfo{year}{2024}\natexlab{}.
\newblock \showarticletitle{Adaptive-{RAG}: Learning to Adapt
  Retrieval-Augmented Large Language Models through Question Complexity}. In
  \bibinfo{booktitle}{\emph{Proceedings of the 2024 Conference of the North
  American Chapter of the Association for Computational Linguistics: Human
  Language Technologies (Volume 1: Long Papers)}}. \bibinfo{pages}{7036--7050}.
\newblock
\href{https://doi.org/10.18653/v1/2024.naacl-long.389}{doi:\nolinkurl{10.18653/v1/2024.naacl-long.389}}


\bibitem[{LangChain AI}(2024)]%
        {langgraph2024}
\bibfield{author}{\bibinfo{person}{{LangChain AI}}.}
  \bibinfo{year}{2024}\natexlab{}.
\newblock \bibinfo{title}{LangGraph: Build Stateful, Multi-Actor Applications
  with {LLM}s}.
\newblock
  \bibinfo{howpublished}{\url{https://langchain-ai.github.io/langgraph/}}.
\newblock
\newblock
\shownote{Accessed June 2026}.


\bibitem[Lewis et~al\mbox{.}(2020)]%
        {lewis2020rag}
\bibfield{author}{\bibinfo{person}{Patrick Lewis}, \bibinfo{person}{Ethan
  Perez}, \bibinfo{person}{Aleksandra Piktus}, \bibinfo{person}{Fabio Petroni},
  \bibinfo{person}{Vladimir Karpukhin}, \bibinfo{person}{Naman Goyal},
  \bibinfo{person}{Heinrich K{\"u}ttler}, \bibinfo{person}{Mike Lewis},
  \bibinfo{person}{Wen-tau Yih}, \bibinfo{person}{Tim Rockt{\"a}schel},
  \bibinfo{person}{Sebastian Riedel}, {and} \bibinfo{person}{Douwe Kiela}.}
  \bibinfo{year}{2020}\natexlab{}.
\newblock \showarticletitle{Retrieval-Augmented Generation for
  Knowledge-Intensive {NLP} Tasks}. In \bibinfo{booktitle}{\emph{Advances in
  Neural Information Processing Systems}}, Vol.~\bibinfo{volume}{33}.
  \bibinfo{pages}{9459--9474}.
\newblock
\urldef\tempurl%
\url{https://proceedings.neurips.cc/paper/2020/hash/6b493230205f780e1bc26945df7481e5-Abstract.html}
\showURL{%
\tempurl}


\bibitem[{Microsoft}(2026a)]%
        {azureHybridSearch2026}
\bibfield{author}{\bibinfo{person}{{Microsoft}}.}
  \bibinfo{year}{2026}\natexlab{a}.
\newblock \bibinfo{title}{{Azure AI Search}: Hybrid Search}.
\newblock
  \bibinfo{howpublished}{\url{https://learn.microsoft.com/en-us/azure/search/hybrid-search-overview}}.
\newblock
\newblock
\shownote{Accessed 2026-06-16}.


\bibitem[{Microsoft}(2026b)]%
        {azureHybridRRF2026}
\bibfield{author}{\bibinfo{person}{{Microsoft}}.}
  \bibinfo{year}{2026}\natexlab{b}.
\newblock \bibinfo{title}{{Azure AI Search}: Reciprocal Rank Fusion}.
\newblock
  \bibinfo{howpublished}{\url{https://learn.microsoft.com/en-us/azure/search/hybrid-search-ranking}}.
\newblock
\newblock
\shownote{Accessed 2026-06-16}.


\bibitem[{Microsoft}(2026c)]%
        {azureSemanticRanker2026}
\bibfield{author}{\bibinfo{person}{{Microsoft}}.}
  \bibinfo{year}{2026}\natexlab{c}.
\newblock \bibinfo{title}{{Azure AI Search}: Semantic Ranking}.
\newblock
  \bibinfo{howpublished}{\url{https://learn.microsoft.com/en-us/azure/search/semantic-search-overview}}.
\newblock
\newblock
\shownote{Accessed 2026-06-16}.


\bibitem[Nogueira and Cho(2019)]%
        {nogueira2019bert}
\bibfield{author}{\bibinfo{person}{Rodrigo Nogueira} {and}
  \bibinfo{person}{Kyunghyun Cho}.} \bibinfo{year}{2019}\natexlab{}.
\newblock \showarticletitle{Passage Re-ranking with {BERT}}.
\newblock \bibinfo{journal}{\emph{arXiv preprint arXiv:1901.04085}}
  (\bibinfo{year}{2019}).
\newblock
\href{https://doi.org/10.48550/arXiv.1901.04085}{doi:\nolinkurl{10.48550/arXiv.1901.04085}}


\bibitem[Qin et~al\mbox{.}(2024)]%
        {qin2023toolllm}
\bibfield{author}{\bibinfo{person}{Yujia Qin}, \bibinfo{person}{Shihao Liang},
  \bibinfo{person}{Yining Ye}, \bibinfo{person}{Kunlun Zhu},
  \bibinfo{person}{Lan Yan}, \bibinfo{person}{Yaxi Lu}, \bibinfo{person}{Yankai
  Lin}, \bibinfo{person}{Xin Cong}, \bibinfo{person}{Xiangru Tang},
  \bibinfo{person}{Bill Qian}, \bibinfo{person}{Sihan Zhao},
  \bibinfo{person}{Lauren Hong}, \bibinfo{person}{Runchu Tian},
  \bibinfo{person}{Ruobing Xie}, \bibinfo{person}{Jie Zhou},
  \bibinfo{person}{Mark Gerstein}, \bibinfo{person}{Dahai Li},
  \bibinfo{person}{Zhiyuan Liu}, {and} \bibinfo{person}{Maosong Sun}.}
  \bibinfo{year}{2024}\natexlab{}.
\newblock \showarticletitle{{ToolLLM}: Facilitating Large Language Models to
  Master 16000+ Real-world APIs}. In \bibinfo{booktitle}{\emph{International
  Conference on Learning Representations}}.
\newblock
\urldef\tempurl%
\url{https://openreview.net/forum?id=dHng2O0Jjr}
\showURL{%
\tempurl}


\bibitem[Reimers and Gurevych(2019)]%
        {reimers2019sentencebert}
\bibfield{author}{\bibinfo{person}{Nils Reimers} {and} \bibinfo{person}{Iryna
  Gurevych}.} \bibinfo{year}{2019}\natexlab{}.
\newblock \showarticletitle{Sentence-{BERT}: Sentence Embeddings using Siamese
  {BERT}-Networks}. In \bibinfo{booktitle}{\emph{Proceedings of the 2019
  Conference on Empirical Methods in Natural Language Processing and the 9th
  International Joint Conference on Natural Language Processing
  (EMNLP-IJCNLP)}}. \bibinfo{pages}{3982--3992}.
\newblock
\href{https://doi.org/10.18653/v1/D19-1410}{doi:\nolinkurl{10.18653/v1/D19-1410}}


\bibitem[Robertson and Zaragoza(2009)]%
        {robertson2009bm25}
\bibfield{author}{\bibinfo{person}{Stephen Robertson} {and}
  \bibinfo{person}{Hugo Zaragoza}.} \bibinfo{year}{2009}\natexlab{}.
\newblock \showarticletitle{The Probabilistic Relevance Framework: {BM25} and
  Beyond}.
\newblock \bibinfo{journal}{\emph{Foundations and Trends in Information
  Retrieval}} \bibinfo{volume}{3}, \bibinfo{number}{4} (\bibinfo{year}{2009}),
  \bibinfo{pages}{333--389}.
\newblock
\href{https://doi.org/10.1561/1500000019}{doi:\nolinkurl{10.1561/1500000019}}


\bibitem[Schick et~al\mbox{.}(2023)]%
        {schick2023toolformer}
\bibfield{author}{\bibinfo{person}{Timo Schick}, \bibinfo{person}{Jane
  Dwivedi-Yu}, \bibinfo{person}{Roberto Dess{\'i}}, \bibinfo{person}{Roberta
  Raileanu}, \bibinfo{person}{Maria Lomeli}, \bibinfo{person}{Eric Hambro},
  \bibinfo{person}{Luke Zettlemoyer}, \bibinfo{person}{Nicolas Cancedda}, {and}
  \bibinfo{person}{Thomas Scialom}.} \bibinfo{year}{2023}\natexlab{}.
\newblock \showarticletitle{Toolformer: Language Models Can Teach Themselves to
  Use Tools}. In \bibinfo{booktitle}{\emph{Advances in Neural Information
  Processing Systems}}, Vol.~\bibinfo{volume}{36}.
\newblock
\urldef\tempurl%
\url{https://arxiv.org/abs/2302.04761}
\showURL{%
\tempurl}


\bibitem[Shinn et~al\mbox{.}(2023)]%
        {shinn2023reflexion}
\bibfield{author}{\bibinfo{person}{Noah Shinn}, \bibinfo{person}{Federico
  Cassano}, \bibinfo{person}{Edward Berman}, \bibinfo{person}{Ashwin Gopinath},
  \bibinfo{person}{Karthik Narasimhan}, {and} \bibinfo{person}{Shunyu Yao}.}
  \bibinfo{year}{2023}\natexlab{}.
\newblock \showarticletitle{Reflexion: Language Agents with Verbal
  Reinforcement Learning}.
\newblock \bibinfo{journal}{\emph{arXiv preprint arXiv:2303.11366}}
  (\bibinfo{year}{2023}).
\newblock
\href{https://doi.org/10.48550/arXiv.2303.11366}{doi:\nolinkurl{10.48550/arXiv.2303.11366}}


\bibitem[Singh et~al\mbox{.}(2025)]%
        {singh2025agenticrag}
\bibfield{author}{\bibinfo{person}{Aditi Singh}, \bibinfo{person}{Abul
  Ehtesham}, \bibinfo{person}{Saket Kumar}, \bibinfo{person}{Tala
  Talaei~Khoei}, {and} \bibinfo{person}{Athanasios~V. Vasilakos}.}
  \bibinfo{year}{2025}\natexlab{}.
\newblock \showarticletitle{Agentic Retrieval-Augmented Generation: A Survey on
  Agentic {RAG}}.
\newblock \bibinfo{journal}{\emph{arXiv preprint arXiv:2501.09136}}
  (\bibinfo{year}{2025}).
\newblock
\href{https://doi.org/10.48550/arXiv.2501.09136}{doi:\nolinkurl{10.48550/arXiv.2501.09136}}


\bibitem[Sun et~al\mbox{.}(2024)]%
        {sun2024dfarag}
\bibfield{author}{\bibinfo{person}{Yiyou Sun}, \bibinfo{person}{Junjie Hu},
  \bibinfo{person}{Wei Cheng}, {and} \bibinfo{person}{Haifeng Chen}.}
  \bibinfo{year}{2024}\natexlab{}.
\newblock \showarticletitle{{DFA}-{RAG}: Conversational Semantic Router for
  Large Language Model with Definite Finite Automaton}. In
  \bibinfo{booktitle}{\emph{Proceedings of the 41st International Conference on
  Machine Learning}} \emph{(\bibinfo{series}{Proceedings of Machine Learning
  Research}, Vol.~\bibinfo{volume}{235})}. \bibinfo{publisher}{PMLR},
  \bibinfo{pages}{47033--47055}.
\newblock
\urldef\tempurl%
\url{https://proceedings.mlr.press/v235/sun24e.html}
\showURL{%
\tempurl}


\bibitem[Thakur et~al\mbox{.}(2021)]%
        {thakur2021beir}
\bibfield{author}{\bibinfo{person}{Nandan Thakur}, \bibinfo{person}{Nils
  Reimers}, \bibinfo{person}{Andreas R{\"u}ckl{\'e}}, \bibinfo{person}{Abhishek
  Srivastava}, {and} \bibinfo{person}{Iryna Gurevych}.}
  \bibinfo{year}{2021}\natexlab{}.
\newblock \showarticletitle{{BEIR}: A Heterogeneous Benchmark for Zero-shot
  Evaluation of Information Retrieval Models}. In
  \bibinfo{booktitle}{\emph{Proceedings of the Neural Information Processing
  Systems Track on Datasets and Benchmarks}}, Vol.~\bibinfo{volume}{1}.
\newblock
\urldef\tempurl%
\url{https://openreview.net/forum?id=wCu6T5xFjeJ}
\showURL{%
\tempurl}


\bibitem[Wen et~al\mbox{.}(2017)]%
        {wen2017network}
\bibfield{author}{\bibinfo{person}{Tsung-Hsien Wen}, \bibinfo{person}{David
  Vandyke}, \bibinfo{person}{Nikola Mrk{\v{s}}i{\'c}}, \bibinfo{person}{Milica
  Ga{\v{s}}i{\'c}}, \bibinfo{person}{Lina~M. Rojas-Barahona},
  \bibinfo{person}{Pei-Hao Su}, \bibinfo{person}{Stefan Ultes}, {and}
  \bibinfo{person}{Steve Young}.} \bibinfo{year}{2017}\natexlab{}.
\newblock \showarticletitle{A Network-based End-to-End Trainable Task-Oriented
  Dialogue System}. In \bibinfo{booktitle}{\emph{Proceedings of the 15th
  Conference of the European Chapter of the Association for Computational
  Linguistics: Volume 1, Long Papers}}. \bibinfo{pages}{438--449}.
\newblock
\urldef\tempurl%
\url{https://aclanthology.org/E17-1042/}
\showURL{%
\tempurl}


\bibitem[Xiao et~al\mbox{.}(2024)]%
        {xiao2024bge}
\bibfield{author}{\bibinfo{person}{Shitao Xiao}, \bibinfo{person}{Zheng Liu},
  \bibinfo{person}{Peitian Zhang}, \bibinfo{person}{Niklas Muennighoff},
  \bibinfo{person}{Defu Lian}, {and} \bibinfo{person}{Jian-Yun Nie}.}
  \bibinfo{year}{2024}\natexlab{}.
\newblock \showarticletitle{C-Pack: Packed Resources For General Chinese
  Embeddings}. In \bibinfo{booktitle}{\emph{Proceedings of the 47th
  International ACM SIGIR Conference on Research and Development in Information
  Retrieval}}. \bibinfo{pages}{641--649}.
\newblock
\href{https://doi.org/10.1145/3626772.3657878}{doi:\nolinkurl{10.1145/3626772.3657878}}
\newblock
\shownote{Accessed June 2026}.


\bibitem[Xu et~al\mbox{.}(2024)]%
        {xu2024cskg}
\bibfield{author}{\bibinfo{person}{Zhentao Xu}, \bibinfo{person}{Mark~Jerome
  Cruz}, \bibinfo{person}{Matthew Guevara}, \bibinfo{person}{Tie Wang},
  \bibinfo{person}{Manasi Deshpande}, \bibinfo{person}{Xiaofeng Wang}, {and}
  \bibinfo{person}{Zheng Li}.} \bibinfo{year}{2024}\natexlab{}.
\newblock \showarticletitle{Retrieval-Augmented Generation with Knowledge
  Graphs for Customer Service Question Answering}. In
  \bibinfo{booktitle}{\emph{Proceedings of the 47th International ACM SIGIR
  Conference on Research and Development in Information Retrieval}}.
  \bibinfo{pages}{2905--2909}.
\newblock
\href{https://doi.org/10.1145/3626772.3661370}{doi:\nolinkurl{10.1145/3626772.3661370}}


\bibitem[Yan et~al\mbox{.}(2024)]%
        {yan2024crag}
\bibfield{author}{\bibinfo{person}{Shi-Qi Yan}, \bibinfo{person}{Jia-Chen Gu},
  \bibinfo{person}{Yun Zhu}, {and} \bibinfo{person}{Zhen-Hua Ling}.}
  \bibinfo{year}{2024}\natexlab{}.
\newblock \showarticletitle{Corrective Retrieval Augmented Generation}.
\newblock \bibinfo{journal}{\emph{arXiv preprint arXiv:2401.15884}}
  (\bibinfo{year}{2024}).
\newblock
\href{https://doi.org/10.48550/arXiv.2401.15884}{doi:\nolinkurl{10.48550/arXiv.2401.15884}}


\bibitem[Yao et~al\mbox{.}(2023)]%
        {yao2023react}
\bibfield{author}{\bibinfo{person}{Shunyu Yao}, \bibinfo{person}{Jeffrey Zhao},
  \bibinfo{person}{Dian Yu}, \bibinfo{person}{Nan Du}, \bibinfo{person}{Izhak
  Shafran}, \bibinfo{person}{Karthik Narasimhan}, {and} \bibinfo{person}{Yuan
  Cao}.} \bibinfo{year}{2023}\natexlab{}.
\newblock \showarticletitle{{ReAct}: Synergizing Reasoning and Acting in
  Language Models}. In \bibinfo{booktitle}{\emph{International Conference on
  Learning Representations}}.
\newblock
\urldef\tempurl%
\url{https://openreview.net/forum?id=WE_vluYUL-X}
\showURL{%
\tempurl}


\bibitem[Young et~al\mbox{.}(2013)]%
        {young2013pomdp}
\bibfield{author}{\bibinfo{person}{Steve Young}, \bibinfo{person}{Milica
  Ga{\v{s}}i{\'c}}, \bibinfo{person}{Blaise Thomson}, {and}
  \bibinfo{person}{Jason~D. Williams}.} \bibinfo{year}{2013}\natexlab{}.
\newblock \showarticletitle{{POMDP}-Based Statistical Spoken Dialog Systems: A
  Review}.
\newblock \bibinfo{journal}{\emph{Proc. IEEE}} \bibinfo{volume}{101},
  \bibinfo{number}{5} (\bibinfo{year}{2013}), \bibinfo{pages}{1160--1179}.
\newblock
\href{https://doi.org/10.1109/JPROC.2012.2225812}{doi:\nolinkurl{10.1109/JPROC.2012.2225812}}


\bibitem[Zhao et~al\mbox{.}(2025)]%
        {zhao2025agentloop}
\bibfield{author}{\bibinfo{person}{Cen Zhao}, \bibinfo{person}{Tiantian Zhang},
  \bibinfo{person}{Hanchen Su}, \bibinfo{person}{Yufeng Zhang},
  \bibinfo{person}{Shaowei Su}, \bibinfo{person}{Mingzhi Xu},
  \bibinfo{person}{Yu Liu}, \bibinfo{person}{Wei Han}, \bibinfo{person}{Jeremy
  Werner}, \bibinfo{person}{Claire~Na Cheng}, {and} \bibinfo{person}{Yashar
  Mehdad}.} \bibinfo{year}{2025}\natexlab{}.
\newblock \showarticletitle{Agent-in-the-Loop: A Data Flywheel for Continuous
  Improvement in {LLM}-based Customer Support}. In
  \bibinfo{booktitle}{\emph{Proceedings of the 2025 Conference on Empirical
  Methods in Natural Language Processing: Industry Track}}.
  \bibinfo{publisher}{Association for Computational Linguistics},
  \bibinfo{address}{Suzhou, China}, \bibinfo{pages}{1919--1930}.
\newblock
\href{https://doi.org/10.18653/v1/2025.emnlp-industry.135}{doi:\nolinkurl{10.18653/v1/2025.emnlp-industry.135}}


\bibitem[Zhou et~al\mbox{.}(2026)]%
        {zhou2026externalization}
\bibfield{author}{\bibinfo{person}{Chenyu Zhou}, \bibinfo{person}{Huacan Chai},
  \bibinfo{person}{Wenteng Chen}, \bibinfo{person}{Zihan Guo},
  \bibinfo{person}{Rong Shan}, \bibinfo{person}{Yuanyi Song},
  \bibinfo{person}{Tianyi Xu}, \bibinfo{person}{Yingxuan Yang},
  \bibinfo{person}{Aofan Yu}, \bibinfo{person}{Weiming Zhang},
  \bibinfo{person}{Congming Zheng}, \bibinfo{person}{Jiachen Zhu},
  \bibinfo{person}{Zeyu Zheng}, \bibinfo{person}{Zhuosheng Zhang},
  \bibinfo{person}{Xingyu Lou}, \bibinfo{person}{Changwang Zhang},
  \bibinfo{person}{Zhihui Fu}, \bibinfo{person}{Jun Wang},
  \bibinfo{person}{Weiwen Liu}, \bibinfo{person}{Jianghao Lin}, {and}
  \bibinfo{person}{Weinan Zhang}.} \bibinfo{year}{2026}\natexlab{}.
\newblock \showarticletitle{Externalization in {LLM} Agents: A Unified Review
  of Memory, Skills, Protocols and Harness Engineering}.
\newblock \bibinfo{journal}{\emph{arXiv preprint arXiv:2604.08224}}
  (\bibinfo{year}{2026}).
\newblock
\href{https://doi.org/10.48550/arXiv.2604.08224}{doi:\nolinkurl{10.48550/arXiv.2604.08224}}


\end{thebibliography}
